\documentclass{JHEP}

\usepackage{amssymb,epsfig}

\newcommand{\tfrac}[2]{{\frac{#1}{#2}}}
\newcommand{\0}[1]{\mathbf{0}^#1}
\newcommand{\1}[1]{\mathbf{1}^#1}
\newcommand{\2}[1]{\mathbf{2}^#1}
\newcommand{\3}[1]{\mathbf{3}^#1}
\newcommand{\4}[1]{\mathbf{4}^#1}
\newcommand{\5}[1]{\mathbf{5}^#1}
\newcommand{\6}[1]{\mathbf{6}^#1}
\newcommand{\7}[1]{\mathbf{7}^#1}
\newcommand{\8}[1]{\mathbf{8}^#1}

\title{Calculating three-loop diagrams
in heavy quark effective theory
with integration-by-parts recurrence relations}

\author{Andrey~G.~Grozin\\
	Institut f\"ur Theoretische Teilchenphysik\\
	Universit\"at Karlsruhe\\
	E-mail: \email{grozin@particle.uni-karlsruhe.de}}

\abstract{An algorithm for calculation of three-loop propagator
diagrams in HQET, based on integration-by-parts recurrence relations,
is constructed and implemented as a \textsf{REDUCE} package
\textsf{Grinder}, and in \textsf{Axiom}.}


\begin{document}

\vspace*{-1.4em}
\section{Introduction}\label{Intro}
\vspace*{-.4em}

Perturbative quantum field theory is progressing fast.  New
high-precision experiments require calculation of higher radiative
corrections --- multiloop Feynman diagrams.  Recently, some
calculations have been done which would seem impossible only a few
years ago.  This is due to the high degree of automation of the
process of generation, analyses and calculation of Feynman diagrams,
which is achieved via extensive use of computer algebra (see~\cite{HS}
for review and references).  All such calculations are performed in
the framework of dimensional regularization~\cite{HV}, i.e.\ diagrams
are calculated as analytical functions of the space-time dimension
$d=4-2\epsilon$.

The integration-by-parts method~\cite{CT} was invented for calculation
of three-loop massless propagator diagrams.  It is the most systematic
method of those currently used, and the most appropriate for
computer-algebra implementation.  It was first implemented as a
\textsf{SCHOONSCHIP}~\cite{V} package \textsf{MINCER}~\cite{GLST}, and
later re-implemented~\cite{LTV} in \textsf{FORM}~\cite{F} (in fact,
\textsf{FORM} was created mainly to run \textsf{MINCER}).  Since then,
\textsf{MINCER} has been the engine behind most of spectacular
successes of perturbative field theory.  Some of these calculations,
with gigabyte-size intermediate expressions, are among the largest
computer-algebra calculations ever undertaken.

Integration-by-parts was used for other classes of problems, too.
Many interesting physical results have been obtained with packages for
\pagebreak[3]
calculating two-loop on-shell massive diagrams and three-loop vacuum
diagrams with a single mass (see, e.g.,~\cite{B,FT,CHKS}).  Reduction
of two-loop propagator integrals with generic masses and momentum to a
finite set of bases integrals has been achieved~\cite{tarasov}.  First
three-loop on-shell calculations have been done recently~\cite{LR,MR}.

Several years ago, an interesting new approach to heavy-quark problems
in Quantum Chromodynamics has been formulated --- Heavy Quark
Effective Theory (HQET), see, e.g., \cite{N,I} for review and
references.  In collaboration with David Broadhurst, I applied the
integration-by-parts method for calculating two-loop propagator
diagrams in HQET~\cite{BG}.  Since then, the algorithm suggested was
used in a large number of physics applications.  A short review of the
integration-by-parts method as applied in heavy quark physics is
presented in~\cite{BG2}.

In the present work, I apply this method for calculating three-loop
propagator diagrams in HQET.  Three-loop anomalous dimensions and
spectral densities in HQET are necessary for a number of physics
applications, such as improved extraction of the $B$ meson decay
constant $f_B$ from lattice simulations and from QCD sum rules.  HQET
lagrangian does not involve mass in the leading order, and, therefore,
the problem is quite similar to the massless one.  My aim is to
produce a reliable package for three-loop HQET calculations, which
could play the same role as \textsf{MINCER} in massless theories.  I
call it \textsf{Grinder}.  Some complication comes from the fact that
there are two kinds of lines now --- massless propagators and
infinitely-heavy ones, and hence the number of diagram topologies is
substantially larger.

In section~\ref{Easy}, we consider three-loop HQET propagator diagrams
with one- or two-loop massless or HQET propagator subdiagrams.  To
this end, we first recall well-known results for massless~\cite{CT}
and HQET~\cite{BG} one- and two-loop diagrams.  After that, diagrams
with two-loop insertions, or with two one-loop insertions, are easily
calculated.  Two-loop HQET diagrams with a single one-loop insertion
are dealt with in a manner similar to the plain two-loop diagrams.  In
section~\ref{Hard}, we consider proper three-loop HQET propagator
diagrams.  Some details of implementation and testing are presented in
section~\ref{Imp}.
\vspace{-.6em}

\section{Diagrams with lower-loop propagator insertions}
\label{Easy}
\vspace{-.4em}

\subsection{Two-loop massless propagator diagrams}\label{Q2}

{\renewcommand\belowcaptionskip{-2em}
\FIGURE{\begin{picture}(32,20)%
\put(16,10){\makebox(0,0){\epsfig{file=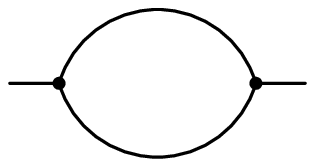}}}
\put(16,0){\makebox(0,0){1}}
\put(16,20){\makebox(0,0){2}}
\end{picture}%
\caption{One-loop massless propagator diagram.\label{q1}}}}
\vspace{-.2em}

The one-loop massless propagator integral (figure~\ref{q1}) can be
easily calculated by using the Feynman parameterization or Fourier
transform to the coordinate space and back:
\begin{eqnarray}
\int \frac{d^d k}{D_1^{n_1} D_2^{n_2}}
&=& i \pi^{d/2} (-p^2)^{d/2-n_1-n_2} G(n_1,n_2)\,,
\nonumber\\
D_1 &=& -k^2\,,\qquad
D_2 = -(k+p)^2\,,
\label{G1}\\\nonumber
G(n_1,n_2) &=& \frac{\Gamma(n_1+n_2-d/2)\Gamma(d/2-n_1)\Gamma(d/2-n_2)}
{\Gamma(n_1)\Gamma(n_2)\Gamma(d-n_1-n_2)}\,.
\end{eqnarray}
The integral with a numerator can be written as a finite sum~\cite{CT}
\begin{eqnarray}
\int \frac{P_n(k) d^d k}{D_1^{n_1} D_2^{n_2}} &=&
 i \pi^{d/2} (-p^2)^{d/2-n_1-n_2} \times
\nonumber\\&&
\times\sum_m G(n_1,n_2;n,m) \frac{(-p^2)^m}{m!}
\left.\left(-\frac{1}{4}\frac{\partial}{\partial k_\mu}
\frac{\partial}{\partial k^\mu}\right)^m
P_n(k) \right|_{k\to p}\,,
\label{G1m}\\
G(n_1,n_2;n,m) &=&\frac{\Gamma(n_1+n_2-m-d/2)
\Gamma(d/2-n_1+n-m)\Gamma(d/2-n_2+m)}
{\Gamma(n_1)\Gamma(n_2)\Gamma(d-n_1-n_2+n)}\,,
\nonumber
\end{eqnarray}
where $P_n(k)$ is an arbitrary homogeneous polynomial: $P_n(\lambda k)=\lambda^n P_n(k)$.

We write the two-loop propagator integral (figure~\ref{q2}\emph{a}) as
\begin{eqnarray}
&&\int \frac{d^d k_1 d^d k_2}{D_1^{n_1} D_2^{n_2} D_3^{n_3} D_4^{n_4} D_5^{n_5}}
= - \pi^d (-p^2)^{d-\sum n_i} G(n_1,n_2,n_3,n_4,n_5)\,,
\nonumber\\
&&D_1 = -k_1^2\,,\qquad
D_2 = -k_2^2\,,\qquad
D_3 = -(k_1+p)^2\,,\qquad
D_4 = -(k_2+p)^2\,,
\nonumber\\
&&D_5 = -(k_1-k_2)^2\,.
\label{G2}
\end{eqnarray}
It is symmetric with respect to $1\leftrightarrow2$,
$3\leftrightarrow4$, and also $1\leftrightarrow3$,
$2\leftrightarrow4$.  If one of the indices is zero, it can be easily
calculated using~(\ref{G1}) (figure~\ref{q2}\emph{b},\emph{c})
\begin{eqnarray}
&&G(n_1,n_2,n_3,n_4,0) = G(n_1,n_3) G(n_2,n_4)\,,
\label{G20a}\\
&&G(0,n_2,n_3,n_4,n_5) = G(n_3,n_5) G\left(n_2,n_4+n_3+n_5- d/2\right)
\label{G20b}
\end{eqnarray}
(and symmetric relations).

\FIGURE[t]{
\begin{picture}(112,22)
\put(56,12){\makebox(0,0){\epsfig{file=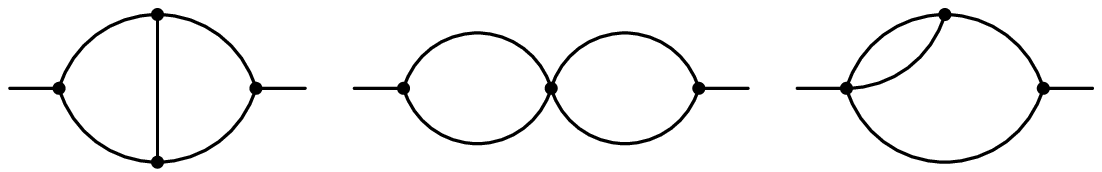}}}
\put(10,3){\makebox(0,0){1}}
\put(22,3){\makebox(0,0){2}}
\put(10,21){\makebox(0,0){3}}
\put(22,21){\makebox(0,0){4}}
\put(18,12){\makebox(0,0){5}}
\put(48.5,4){\makebox(0,0){1}}
\put(63.5,4){\makebox(0,0){2}}
\put(48.5,20){\makebox(0,0){3}}
\put(63.5,20){\makebox(0,0){4}}
\put(96,2){\makebox(0,0){2}}
\put(90,21){\makebox(0,0){3}}
\put(102,21){\makebox(0,0){4}}
\put(95,13){\makebox(0,0){5}}
\put(16,-3){\makebox(0,0)[b]{$a$}}
\put(56,-3){\makebox(0,0)[b]{$b$}}
\put(96,-3){\makebox(0,0)[b]{$c$}}
\end{picture}
\caption{Two-loop massless propagator diagram}
\label{q2}}

Applying the operators $\partial_1\cdot(k_1-k_2)$ and $\partial_1\cdot k_1$
(where $\partial_i={\partial}/{\partial k_i}$)
to the integrand of~(\ref{G2}),
we obtain the recurrence relations for $G(n_1,n_2,n_3,n_4,n_5)$
(known as triangle relations~\cite{CT})
\begin{eqnarray}
\left[ d-n_1-n_3-2n_5 + n_1\1+(\2--\5-) + n_3\3+(\4--\5-) \right] G &=& 0\,,
\label{t1}\\
\left[ d-n_3-n_5-2n_1 + n_3\3+(1  -\1-) + n_5\5+(\2--\1-) \right] G &=& 0\,,
\label{t2}
\end{eqnarray}
where, for example,
\begin{equation}
\1\pm G(n_1,n_2,n_3,n_4,n_5)=G(n_1\pm1,n_2,n_3,n_4,n_5)\,.
\label{ladop}
\end{equation}

\pagebreak[3]

\noindent Of course, more relations are obtained by symmetry.  Another
interesting relation is derived by applying the operator
$\frac{\partial}{\partial p}\cdot(k_2+p)$.  Substituting the general
form of the relevant vector integral, we arrive at
\begin{eqnarray}
\Biggl[ \tfrac{1}{2}d+n_4-n_1-n_2-n_5
+ \left(\tfrac{3}{2}d-n_1-n_2-n_3-n_4-n_5\right)(\4--\2-)+
\hphantom{\Biggr] G}&&
\label{t3}\\\nonumber
 + n_3\3+(\4--\5-) \Biggr] G &=& 0\,.
\end{eqnarray}
This formula was derived long ago by S.A.~Larin in his M.Sc.\ thesis~\cite{L}
(again, similar relations follow by symmetries).

If indices of two adjacent lines are non-positive integers, the
integral contains a no-scale vacuum subdiagram and hence vanishes.
The cases with zero indices are given by~(\ref{G20a}), (\ref{G20b}).
When $n_5<0$ and $n_3\ne1$, $n_5$ can be raised by~(\ref{t3}); the
cases $n_1\ne1$, $n_2\ne1$, $n_4\ne1$ are symmetric.  The case
$n_5<0$, $n_1=n_1=n_3=n_4=1$ is handled by
\begin{eqnarray}
\left[ (d-2n_5-4)\5+ + 2(d-n_5-3) \right] G(1,1,1,1,n_5)=\qquad\qquad
\nonumber\\
= 2\1+(\3--\2-\5+) G(1,1,1,1,n_5)\,,
\label{r5}
\end{eqnarray}
which follows from~(\ref{t1}) and~(\ref{t2}) at $n_1=n_2=n_3=n_4=1$ (note
that the terms in the right-hand side of~(\ref{r5}) are trivial for
any $n_5$; for $n_5<0$, they vanish).  When $n_2<0$, it can be raised
by~(\ref{t3}); the cases $n_1<0$, $n_3<0$, $n_4<0$ are\linebreak symmetric.

\FIGURE[t]{
\centerline{\begin{picture}(77,22)
\put(38.5,12){\makebox(0,0){\epsfig{file=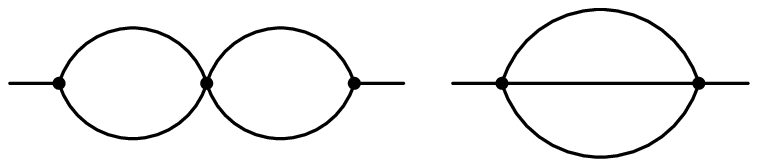}}}
\put(21,-3){\makebox(0,0)[b]{$a$}}
\put(61,-3){\makebox(0,0)[b]{$b$}}
\end{picture}}
\caption{Basis two-loop massless propagator integrals}
\label{qb2}}
We are left with the most important situation when all the indices are
positive.  Applying~(\ref{t1}), we reduce $n_2$, $n_4$, $n_5$ until
one of them vanishes.  Then~(\ref{G20a}) and~(\ref{G20b}) apply.  If
$\max(n_1,n_3)<\max(n_2,n_4)$, it is more efficient to lower $n_1$,
$n_3$, $n_5$.

All one-loop integrals (figure~\ref{q1}) with integer $n_{1,2}$ are
proportional to $G_1=G(1,1)$, the coefficient being a rational
function of $d$.  All two-loop integrals with integer indices reduce
to $G_1^2$ (figure~\ref{qb2}\emph{a}) and $G_2=G(0,1,1,0,1)$
(figure~\ref{qb2}\emph{b}), with rational coefficients.  Here
\begin{equation}
G_n = \frac{1}{\left(n+1-n\frac{d}{2}\right)_n
\left((n+1)\frac{d}{2}-2n-1\right)_n}
\frac{\Gamma(1+n\epsilon)\Gamma^{n+1}
(1-\epsilon)}{\Gamma(1-(n+1)\epsilon)}\,,
\label{qb}
\end{equation}
where $(x)_n=\Gamma(x+n)/\Gamma(x)$ is the Pochhammer symbol.

\pagebreak[3]

\subsection{Two-loop HQET propagator diagrams}
\label{H2}

\FIGURE{\begin{picture}(32,12)%
\put(16,6){\makebox(0,0){\epsfig{file=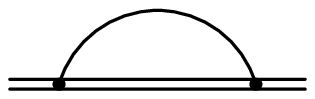}}}
\put(16,-1){\makebox(0,0){1}}
\put(16,13){\makebox(0,0){2}}
\end{picture}%
\caption{One-loop HQET propagator diagram.\label{h1}}}

The one-loop HQET propagator integral (figure~\ref{h1})
can be easily calculated by using the modified Feynman parameterization
(see, e.g., \cite{BG,BG2}), or Fourier transform to the coordinate space and back:
\begin{eqnarray}
\int \frac{d^d k}{D_1^{n_1} D_2^{n_2}}
&=& i \pi^{d/2} (-2\omega)^{d-2n_2} I(n_1,n_2)\,,
\nonumber\\
D_1 &=& \frac{(k+p)\cdot v}{\omega}\,,\qquad
D_2 = -k^2\,,
\nonumber\\
I(n_1,n_2) &=& \frac{\Gamma(n_1+2n_2-d)
\Gamma(d/2-n_2)}{\Gamma(n_1)\Gamma(n_2)}
\label{I1}
\end{eqnarray}
(here $v$ is 4-velocity of the heavy quark, $v^2=1$, and $\omega =
p\cdot v$ is the residual energy).  Similarly to~(\ref{G1m}), we
obtain
\begin{eqnarray}
\int \frac{P_n(k) d^d k}{D_1^{n_1} D_2^{n_2}} &=&
 i \pi^{d/2} (-2\omega)^{d-2n_2}\times
\nonumber\\&&
\times \sum_m I(n_1,n_2;n,m) \frac{(-2\omega)^{2m}}{m!}
\left.\left(-\frac{1}{4}\frac{\partial}{\partial k_\mu}
\frac{\partial}{\partial k^\mu}\right)^m
P_n(k) \right|_{k\to 2\omega v}\,,
\nonumber\\
I(n_1,n_2;n,m) &=& \frac{\Gamma(n_1+2n_2-n-d)
\Gamma(d/2-n_2+n-m)}{\Gamma(n_1)\Gamma(n_2)}\,.
\label{I1m}
\end{eqnarray}

There are two topologies of two-loop propagator HQET diagrams
(figure~\ref{h2}\emph{a},\emph{b}).  We write the first of them as
\begin{equation}
\begin{array}[b]{rclcrcl}
\displaystyle\int \frac{d^d k_1 d^d k_2}{D_1^{n_1} D_2^{n_2} D_3^{n_3} D_4^{n_4} D_5^{n_5}}
&=& \multicolumn{5}{l}{- \pi^d (-2\omega)^{2(d-n_3-n_4-n_5)} I(n_1,n_2,n_3,n_4,n_5)\,,}
\\
D_1 &=&\displaystyle \frac{(k_1+p)\cdot v}{\omega}\,,&\qquad&
D_2 &=&\displaystyle  \frac {(k_2+p)\cdot v}\omega \,,
\\
D_3 &=& -k_1^2\,,&\qquad&
D_4 &=& -k_2^2\,,\qquad
D_5 = -(k_1-k_2)^2\,.
\label{I2}
\end{array}
\end{equation}
It is symmetric with respect to $1\leftrightarrow3$, $2\leftrightarrow4$.
If one of the indices is zero, it can be easily calculated
using~(\ref{I1})  and~(\ref{G1})
(figure~\ref{h2}\emph{c},\emph{d},\emph{e})
\begin{eqnarray}
I(n_1,n_2,n_3,n_4,0) &=& I(n_1,n_3) I(n_2,n_4)\,,
\label{I20a}\\
I(0,n_2,n_3,n_4,n_5) &=& G(n_3,n_5) I(n_2,n_4+n_3+n_5-d/2)\,,
\label{I20b}\\
I(n_1,n_2,0,n_4,n_5) &=& I(n_1,n_5) I(n_2+n_1+2n_5-d,n_4)
\label{I20c}
\end{eqnarray}
(and symmetric relations).

\FIGURE[t]{
\begin{picture}(112,45)
\put(56,24){\makebox(0,0){\epsfig{file=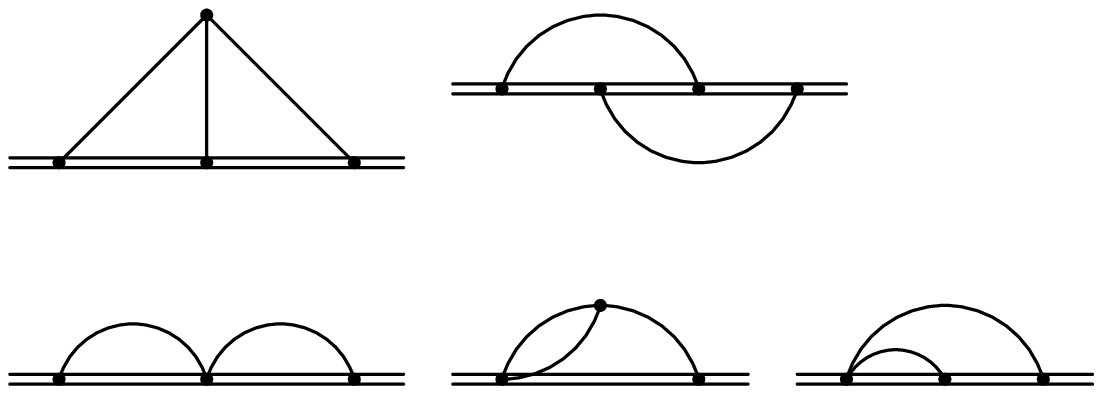}}}
\put(21,19){\makebox(0,0)[b]{$a$}}
\put(13.5,24){\makebox(0,0){1}}
\put(28.5,24){\makebox(0,0){2}}
\put(9,34){\makebox(0,0){3}}
\put(33,34){\makebox(0,0){4}}
\put(23,34){\makebox(0,0){5}}
\put(66,19){\makebox(0,0)[b]{$b$}}
\put(56,32){\makebox(0,0){1}}
\put(66,32){\makebox(0,0){3}}
\put(76,38){\makebox(0,0){2}}
\put(61,45){\makebox(0,0){4}}
\put(71,25){\makebox(0,0){5}}
\put(21,-3){\makebox(0,0)[b]{$c$}}
\put(13.5,2.5){\makebox(0,0){1}}
\put(28.5,2.5){\makebox(0,0){2}}
\put(13.5,14){\makebox(0,0){3}}
\put(28.5,14){\makebox(0,0){4}}
\put(61,-3){\makebox(0,0)[b]{$d$}}
\put(61,2.5){\makebox(0,0){2}}
\put(54,13){\makebox(0,0){3}}
\put(68,13){\makebox(0,0){4}}
\put(61,9){\makebox(0,0){5}}
\put(96,-3){\makebox(0,0)[b]{$e$}}
\put(91,2.5){\makebox(0,0){1}}
\put(101,2.5){\makebox(0,0){2}}
\put(96,15){\makebox(0,0){4}}
\put(96,9){\makebox(0,0){5}}
\end{picture}
\caption{Two-loop HQET propagator diagram}
\label{h2}}

Applying the operators $\partial_1\cdot(k_1-k_2)$, $\partial_1\cdot k_1$ and
$\partial_1\cdot v$ to the integrand of~(\ref{I2}),
we obtain~\cite{BG}
\begin{eqnarray}
\left[ d-n_1-n_3-2n_5 + n_1\1+\2- + n_3\3+(\4--\5-) \right] I &=& 0\,,
\label{ht1}\\
\left[ d-n_1-n_5-2n_3 + n_1\1+ + n_5\5+(\4--\3-) \right] I &=& 0\,,
\label{ht2}\\
\left[ -2n_1\1+ + n_3\3+(\1--1) + n_5\5+(\1--\2-) \right] I &=& 0\,.
\label{ht3}
\end{eqnarray}
Of course, more relations are obtained by symmetry.  Applying
$\omega\frac{d}{d\omega}$ and using homogeneity in $\omega$, we obtain
\begin{equation}
\left[ 2(d-n_3-n_4-n_5)-n_1-n_2 + n_1\1+ + n_2\2+ \right] I = 0\,,
\label{hom}
\end{equation}
which is nothing but the sum of~(\ref{ht2}) and its mirror-symmetric.
Subtracting the $\2-$ shifted version of~(\ref{hom}) from~(\ref{ht1}),
we obtain the most useful relation~\cite{BG}
\begin{eqnarray}
\Bigl[ d-n_1-n_2-n_3-2n_5+1-\hphantom{\Bigr] I}
\nonumber\\
-\, \bigl(2(d-n_3-n_4-n_5)-n_1-n_2+1\bigr)\2-
+ n_3\3+(\4--\5-) \Bigr] I &=& 0\,,
\label{djb}
\end{eqnarray}
which lowers $n_2$, $n_4$, $n_5$, and does not raise heavy-quark indices.

If indices of two adjacent lines are non-positive integers, the
integral contains a no-scale vacuum subdiagram and hence vanishes.
The cases with zero indices are given by~(\ref{I20a}), (\ref{I20b})
and~(\ref{I20c}).  When $n_2<0$, it can be raised by~(\ref{djb}); the
case $n_1<0$ is symmetric.  Similarly, if $n_3<0$, it can be raised
by~(\ref{djb}); the case $n_4<0$ is symmetric.  When $n_5<0$ and
$n_3\ne1$, $n_5$ can be raised by~(\ref{djb}) (the case $n_4\ne1$ is
symmetric); when $n_5<0$ and $n_1\ne1$, $n_5$ can be raised
by~(\ref{ht2}) (the case $n_2\ne1$ is symmetric).  The case $n_5<0$,
$n_1=n_1=n_3=n_4=1$ is handled by
\begin{eqnarray}
\left[ (d-2n_5-4)\5+ - 2(d-n_5-3) \right] I(1,1,1,1,n_5) =\qquad\qquad\qquad
\nonumber\\
\qquad\qquad= \left[ (2d-2n_5-7)\1-\5+ - \3-\4+\5- + \1-\3+ \right] I(1,1,1,1,n_5) \,,
\label{hr5}
\end{eqnarray}
which follows from~(\ref{ht1}), (\ref{ht2})  and~(\ref{ht3}) at
$n_1=n_2=n_3=n_4=1$ (note that the terms on the right-hand side
of~(\ref{hr5}) are trivial for any $n_5$; for $n_5<0$, they vanish).

We are left with the most important situation when all the indices are
positive.  Applying~(\ref{djb}), we reduce $n_2$, $n_4$, $n_5$ until
one of them vanishes.  Then~(\ref{I20a}), (\ref{I20b})
and~(\ref{I20c}) apply.  If $\max(n_1,n_3)<\max(n_2,n_4)$, it is more
efficient to lower \mbox{$n_1$, $n_3$, $n_5$}.

In the second topology (figure~\ref{h2}\emph{b}), three heavy-quark
denominators depend on only two variables $k_{1,2}\cdot v$, hence they
are linearly dependent.  Therefore, there is one scalar product which
cannot be expressed via the denominators.  Let's define the integral
\begin{equation}
\begin{array}[b]{rclcrclcrcl}
\multicolumn{11}{l}{\displaystyle\int \frac{N^{n_0} d^d k_1 d^d k_2}
{D_1^{n_1} D_2^{n_2} D_3^{n_3} D_4^{n_4} D_5^{n_5}}
= - \pi^d (-2\omega)^{2(d+n_0-n_4-n_5)} J(n_1,n_2,n_3,n_4,n_5;n_0)\,,}
\\[9pt]
 D_1 &=&\displaystyle \frac{(k_1+p)\cdot  v}{\omega}\,,&\qquad&
D_2 &=&\displaystyle \frac{(k_2+p)\cdot v}{\omega}\,,&\qquad&
D_3 &=&\displaystyle \frac{(k_1+k_2+p)\cdot  v}{\omega}\,,
\\
 D_4 &=& -k_1^2\,,&\qquad& 
D_5 &=& -k_2^2\,,&\qquad& 
N &=& 2 k_1\cdot k_2
\end{array}
\label{J2}
\end{equation}
(it is symmetric with respect to $1\leftrightarrow2$, $4\leftrightarrow5$).
Noting that $D_1+D_2-D_3=1$, we immediately have~\cite{BG}
\begin{equation}
(1-\1--\2-+\3-) J = 0\,.
\label{j0}
\end{equation}
Applying $\partial_1\cdot k_1$, $\partial_1\cdot k_2$
and $\partial_1\cdot v$ to the integrand, we have
\begin{eqnarray}
\left[ d+n_0-n_1-n_3-2n_4 + n_1\1+ + n_3\3+\2- \right] J &=& 0
\label{j1}\\
\left[ n_1-n_3 - n_1\1+\3- + n_3\3+\1- + n_4\4+\0+ - 2n_0\0-\5- \right] J &=& 0
\label{j2}\\
\left[ - 2n_1\1+ - 2n_3\3+ + n_4\4+(\1--1) + n_0\0-(\2--1) \right] J &=& 0\,.
\label{j3}
\end{eqnarray}
Homogeneity in $\omega$ gives $[2(d+n_0-n_4-n_5)-n_1-n_2-n_3 + n_1\1+
+ n_2\2+ + n_3\3+] J = 0$, which is nothing but the sum of~(\ref{j1})
and its mirror-symmetric.  The boundary values of the integral
(figure~\ref{h2}\emph{c},\emph{e}) are
\begin{eqnarray}
J(n_1,n_2,0,n_4,n_5;n_0) &=& (\5--\3--\4-)^{n_0} I(n_1,n_2,n_4,n_5,0)
\label{j01}\\
J(0,n_2,n_3,n_4,n_5;n_0) &=& (\4--\3-+\5-)^{n_0} I(n_3,n_2,0,n_5,n_4)
\label{j02}
\end{eqnarray}
and the symmetric relation for $n_2=0$.
If $n_0=0$,
\begin{eqnarray}
J(n_1,n_2,0,n_4,n_5) &=& I(n_1,n_4) I(n_2,n_5)
\label{j03}\\
J(0,n_2,n_3,n_4,n_5) &=& I(n_3,n_4) I(n_2+n_3+2n_4-d,n_5)\,.
\end{eqnarray}
If $n_4\le0$, or $n_5\le0$, or two adjacent heavy-quark indices are
non-positive, the integral vanishes.  If any of $n_1$, $n_2$, $n_3$ is
negative, it can be raised by~(\ref{j0}).  If all of them are
positive, we use~(\ref{j0}) to lower $n_1$, $n_2$ or $n_3$, until one
of these indices vanish.

Instead of using~(\ref{j01}), (\ref{j02}) when $n_0>0$, we could
proceed in another way.  If $n_4>1$, we lower both $n_0$ and $n_4$
using~(\ref{j2}) (the case $n_5>1$ is symmetric).  We are left with
$J(n_1,n_2,0,1,1;n_0)$ and $J(0,n_2,n_3,1,1;n_0)$.  In the first case,
if $n_1>1$, we lower it using~(\ref{j3}) (the case $n_2>1$ is
symmetric).  In the second case, if $n_3>1$, we lower it
using~(\ref{j3}); if $n_2>1$, we lower it using the difference
of~(\ref{j3}) and its mirror-symmetric.  The two remaining cases
(figure~\ref{h2}\emph{c},\emph{e}) are easily evaluated
using~(\ref{I1m}):
\begin{eqnarray*}
J(1,1,0,1,1;n_0) &=& (-1)^{n_0} I(1,1;n_0,0) I(1,1)\,,\\
J(0,1,1,1,1;n_0) &=& I(1,1;n_0,0)
\sum_{l=0}^{n_0} \frac{(-1)^l n_0!}{l! (n_0-l)!} I(4-2n_0+l-d,1)
\end{eqnarray*}

All one-loop integrals (figure~\ref{h1}) with integer $n_{1,2}$ are
proportional to $I_1=I(1,1)$, the coefficient being a rational
function of $d$.  All two-loop integrals with integer indices reduce
to $I_1^2$ (figure~\ref{hb2}\emph{a}) and $I_2=I(0,1,1,0,1)$
(figure~\ref{hb2}\emph{b}), with rational coefficients.  Here
\begin{equation}
I_n = \frac{1}{(1-n(d-2))_{2n}} \Gamma(1+2n\epsilon)\Gamma^n(1-\epsilon)\,.
\label{hb}
\end{equation}

\FIGURE[t]{
\begin{minipage}{\textwidth}
\begin{center}
\begin{picture}(77,13)
\put(38.5,6.5){\makebox(0,0){\epsfig{file=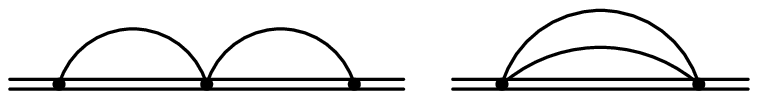}}}
\put(21,-3){\makebox(0,0)[b]{$a$}}
\put(61,-3){\makebox(0,0)[b]{$b$}}
\end{picture}
\end{center}
\end{minipage}
\caption{Basis two-loop HQET propagator integrals}
\label{hb2}}

\subsection{Three-loop HQET diagrams with lower-loop insertions}
\label{Ha}

Diagrams with a two-loop propagator insertion,
or with two one-loop insertions, figure~\ref{i1},
are trivially calculated by multiplying the relevant insertion[s]
by the one-loop HQET integral with non-integer indices,
whose values are obvious by dimensionality.

\FIGURE[t]{
\begin{picture}(87,52)
\put(43.5,26){\makebox(0,0){\epsfig{file=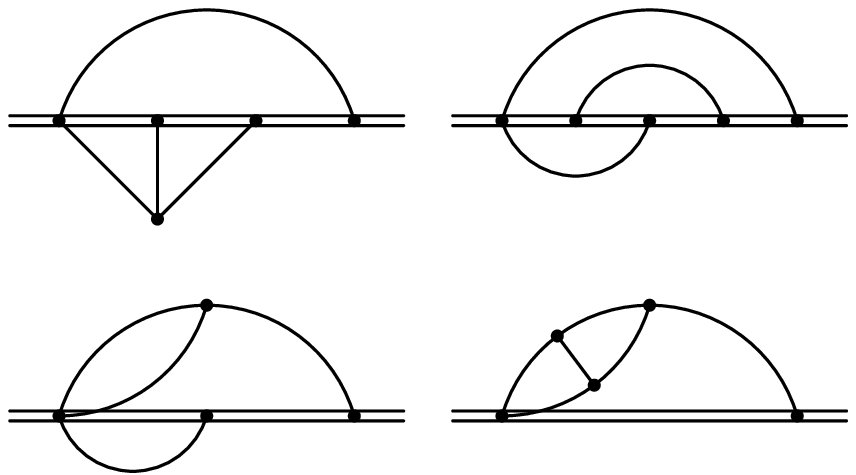}}}
\put(21,25){\makebox(0,0)[b]{$a$}}
\put(66,25){\makebox(0,0)[b]{$b$}}
\put(21,-3){\makebox(0,0)[b]{$c$}}
\put(66,-3){\makebox(0,0)[b]{$d$}}
\end{picture}
\caption{Three-loop HQET propagator diagrams with a two-loop propagator insertion,
or with two one-loop insertions}
\label{i1}}

Next we consider the diagrams obtained from the two-loop HQET diagram of figure~\ref{h2}\emph{a}
by adding a single one-loop propagator insertion (figure~\ref{i2}).
They are equal to the product of the corresponding one-loop integral
and the integral of figure~\ref{h2}\emph{a} with one non-integer index.
All relations derived for the diagram of figure~\ref{h2}\emph{a} are valid;
however, the non-integer index changes the strategy of their application.

\FIGURE[t]{
\begin{picture}(132,24)
\put(66,12){\makebox(0,0){\epsfig{file=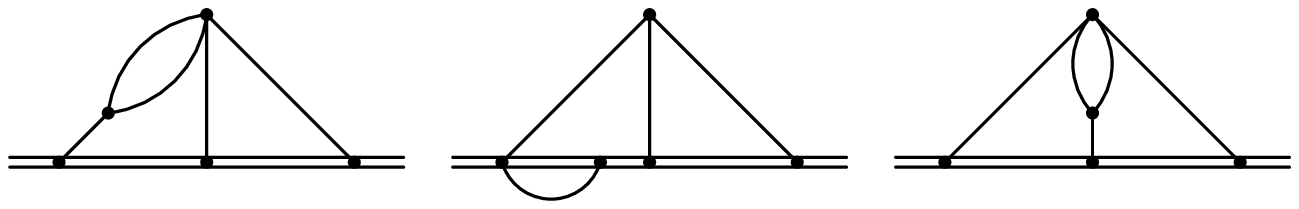}}}
\put(21,-3){\makebox(0,0)[b]{$a$}}
\put(66,-3){\makebox(0,0)[b]{$b$}}
\put(111,-3){\makebox(0,0)[b]{$c$}}
\end{picture}
\caption{The diagram of figure~\protect\ref{h2}\emph{a} with a single one-loop insertion}
\label{i2}}

Let's consider the case of figure~\ref{i2}\emph{a} with a non-integer
$n_3$.  When $n_5<0$, it can always be raised by~(\ref{djb}).  When
$n_2<0$, it can be raised by~(\ref{djb}), too; when $n_1<0$, it can be
raised by the mirror-symmetric relation.  When $n_4<0$, it can always
be raised by~(\ref{djb}), too.  Finally, when all indices $n_1$,
$n_2$, $n_4$, $n_5$ are positive, we use~(\ref{djb}) to lower $n_2$,
$n_4$ or $n_5$, until~(\ref{I20a}), (\ref{I20b}) and~(\ref{I20c}) are
reached.

In the case of figure~\ref{i2}\emph{b}, $n_1$ is non-integer.  When
$n_5<0$ and $n_3\ne1$, we can raise $n_5$ by~(\ref{djb}) (the case
$n_4\ne1$ is symmetric); when $n_5<0$ and $n_2\ne1$, we can lower
$n_2$ by the relation symmetric to~(\ref{ht2}); when $n_5<0$ and
$n_2=n_3=n_4=1$, we can use the relation
\begin{equation}
\left[ 2(d-n_5-3) - \4+ - n_5\5+\1- \right] I(n_1,1,1,1,n_5) = 0\,,
\label{r5111}
\end{equation}
which follows from~(\ref{ht3}) and~(\ref{ht2}), and then the term with
$\4+$ can be treated as above.  When $n_2<0$, it can be raised
by~(\ref{djb}).  When $n_3<0$ and $n_4\ne1$, we can raise $n_3$ by the
relation symmetric to~(\ref{djb}); when $n_3<0$ and $n_5\ne1$, we can
raise $n_3$ by~(\ref{ht2}); when $n_3<0$ and $n_2\ne1$, we lower $n_2$
by~(\ref{hom}); finally, when $n_3<0$ and $n_2=n_4=n_5=1$, we
use~(\ref{djb}) to get a trivial term with $n_2=0$.
When $n_4<0$ and $n_3\ne1$, we can raise $n_4$ by~(\ref{djb}); when
$n_4<0$ and $n_5\ne1$, we can raise $n_4$ by~(\ref{ht2}); when $n_4<0$
and $n_2\ne1$, we lower $n_2$ by~(\ref{hom}); finally, when $n_4<0$
and $n_2=n_3=n_5=1$, we use~(\ref{ht3}) to raise $n_3$ or $n_5$, 
and proceed as above.  Finally, when all indices
$n_2$, $n_3$, $n_4$, $n_5$ are positive, we use~(\ref{djb}) to lower
$n_2$, $n_4$ or $n_5$, until~(\ref{I20a}), (\ref{I20b})
and~(\ref{I20c}) are reached.

In the case of figure~\ref{i2}\emph{c}, $n_5$ is non-integer.  When
$n_2<0$, it can be raised by~(\ref{djb}); the case $n_1<0$ is
symmetric.  When $n_3<0$, it can be raised by~(\ref{ht2}); the case
$n_4<0$ is symmetric.  When $n_1>1$, it can be lowered by~(\ref{ht2});
the case $n_2>1$ is symmetric.  When $n_3>1$, it can be lowered
by~(\ref{ht3}), with a trivial additional term having $n_1=0$; the
case $n_4>1$ is symmetric.  We are left with $n_1=n_2=n_3=n_4=1$; the
relation~(\ref{hr5}) can be used to lower or raise $n_5$, with trivial
additional terms.  An integral with some specific value of $n_5$ (of
the form integer plus $\epsilon$) has to be considered as a new basis
element.  This integral has been calculated, exactly in $d$
dimensions, in~\cite{BB} in terms of ${}_3F_2$ hypergeometric
functions, using the Hegenbauer polynomial technique in the coordinate
space~\cite{CKT}.

Finally, we consider the diagrams obtained from the two-loop HQET
diagram of figure~\ref{h2}\emph{b} by adding a single one-loop
propagator insertion (figure~\ref{i3}).  The diagram of
figure~\ref{i3}\emph{a}, with a non-integer $n_4$ is calculated
exactly as the two-loop one.

\FIGURE[t]{
\begin{picture}(132,20)
\put(66,10){\makebox(0,0){\epsfig{file=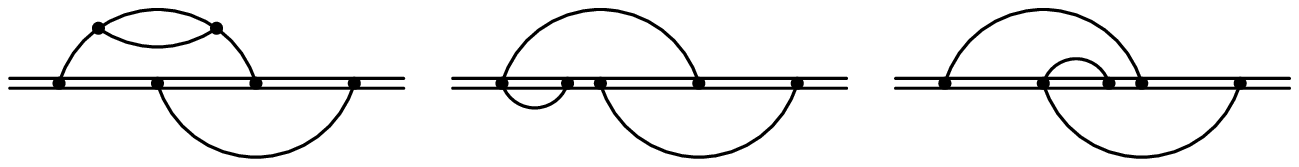}}}
\put(21,-3){\makebox(0,0)[b]{$a$}}
\put(66,-3){\makebox(0,0)[b]{$b$}}
\put(111,-3){\makebox(0,0)[b]{$c$}}
\end{picture}
\caption{The diagram of figure~\protect\ref{h2}\emph{b} with a single one-loop insertion}
\label{i3}}

In the case of figure~\ref{i3}\emph{b}, $n_1$ is non-integer.  When
$n_3<0$ or $n_2<0$, they can be raised by~(\ref{j0}).  In the case of
positive indices, we use
\begin{equation}
\left[ d-n_3-2n_4 + n_1\1+(\2--\3-) + n_3\3+\2- \right] J = 0
\label{j4}
\end{equation}
(which follows from~(\ref{j1})  and~(\ref{j0})).  It either lowers
$n_2+n_3$, or, at a fixed $n_2+n_3$, lowers $n_2$.  Therefore, sooner
or later, we reach~(\ref{j01}) and~(\ref{j02}).

In the case of figure~\ref{i3}\emph{c}, $n_3$ is non-integer.  If
$n_1<0$ or $n_2<0$, they can be raised by~(\ref{j0}).  When $n_4>1$,
we can lower it or $n_1$ by~(\ref{j3}); the case $n_5>1$ is symmetric.
When $n_1>1$, it can be lowered by~(\ref{j1}); the case $n_2>1$ is
symmetric.  We are left with $n_1=n_2=n_4=n_5=1$; the
relation~(\ref{j0}) can be used to lower or raise $n_3$, with trivial
additional terms.  An integral with some specific value of $n_3$ (of
the form integer plus $2\epsilon$) has to be considered as a new basis
element.

It is not difficult to calculate $J(1,1,n_3,n_4,n_5)$ for arbitrary $n_{3,4,5}$
(not necessarily integer) in the coordinate space:
\begin{eqnarray}
J(1,1,n,n_1,n_2) &=& \frac{\Gamma(n-2(d-n_1-n_2-1))\Gamma(d/2-n_1)\Gamma(d/2-n_2)}
{\Gamma(n)\Gamma(n_1)\Gamma(n_2)} J\,,
\nonumber\\
J &=& t^{2(d-n_1-n_2)-n-1} \int\limits_{0<t_1<t_2<t} dt_1\,dt_2\,
t_2^{2n_1-d} (t-t_1)^{2n_2-d} (t_2-t_1)^{n-1}
\nonumber\\[-2pt]
 &=& \frac{1}{n(2n_1+n+1-d)}\,_3F_2\Biggl(
\begin{array}{c}1,\;d-2n_2,\;2n_1+n+1-d\\
n+1,\;2n_1+n+2-d\end{array}\Bigg|\,1\Biggr).\qquad
\label{Jm}
\end{eqnarray}

All diagrams considered in this section are particular cases of the
generic three-loop topologies, which will be discussed in
section~\ref{Hard}, when some lines are shrunk (i.e., some indices
vanish).  Therefore, we don't consider diagrams with numerators here:
numerators should be dealt with in the context of generic topologies,
and the formulae of this section are used only as boundary values for
the corresponding recurrence relations, after elimination of
numerators.

All three-loop HQET propagator integrals with lower-loop propagator
insertions are linear combinations of 7 basis integrals
(figure~\ref{hb3}\emph{a}--\emph{g}), coefficients being rational
functions of $d$.  The basis integrals of figure~\ref{hb3}\emph{a}
($I_1^3$), figure~\ref{hb3}\emph{b} ($I_1 I_2$),
figure~\ref{hb3}\emph{c} ($I_3$), figure~\ref{hb3}\emph{d} ($I_3
I_1^2/I_2$), and figure~\ref{hb3}\emph{e} ($I_3 G_1^2/G_2$) are known
exactly in $d$ dimensions in terms of $\Gamma$ functions.  Those of
figure~\ref{hb3}\emph{f},\emph{g} contain hypergeometric ${}_3 F_2$
functions of the unit argument.  Several terms of their expansion in
$\epsilon$ can be obtained using the methods which were
recently developed in~\cite{Bz}.  As we shall see in
section~\ref{Hard}, there is only one additional basis integral,
figure~\ref{hb3}\emph{h}.

\FIGURE[t]{
\begin{picture}(122,67.5)
\put(61,33.75){\makebox(0,0){\epsfig{file=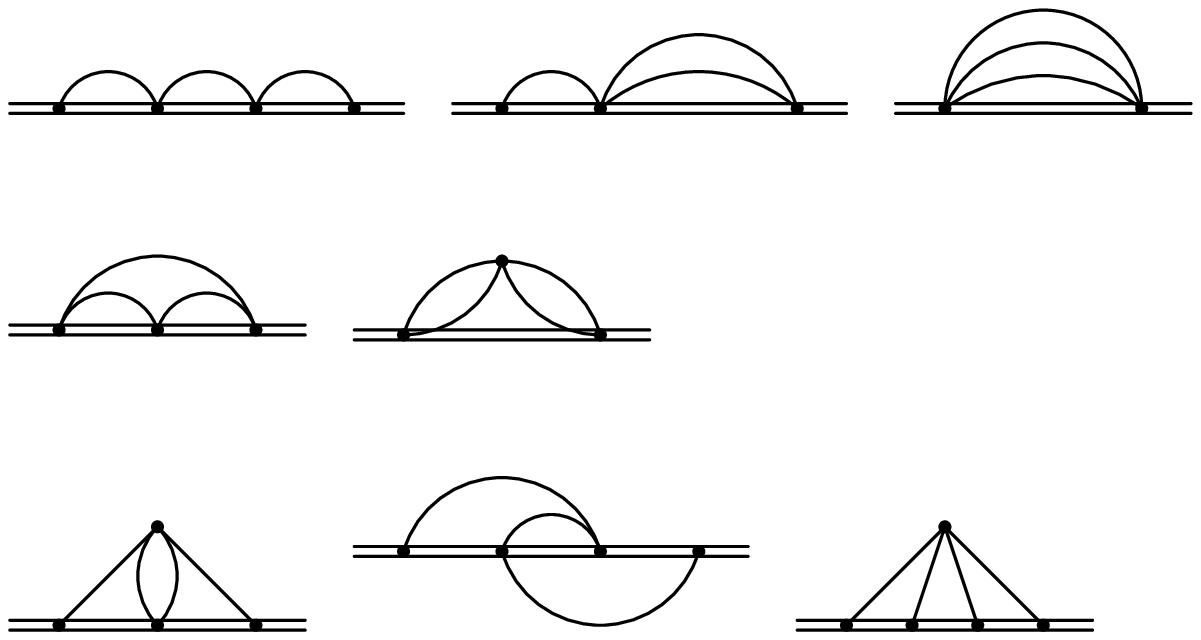}}}
\put(21,49.5){\makebox(0,0)[b]{$a$}}
\put(66,49.5){\makebox(0,0)[b]{$b$}}
\put(106,49.5){\makebox(0,0)[b]{$c$}}
\put(16,27){\makebox(0,0)[b]{$d$}}
\put(51,27){\makebox(0,0)[b]{$e$}}
\put(16,-3){\makebox(0,0)[b]{$f$}}
\put(56,-3){\makebox(0,0)[b]{$g$}}
\put(96,-3){\makebox(0,0)[b]{$h$}}
\end{picture}
\caption{Basis tree-loop HQET propagator integrals}
\label{hb3}}

\section{Proper three-loop HQET propagator diagrams}
\label{Hard}

In the massless case~\cite{CT}, there are only 3 topologies of proper
three-loop propagator diagrams: Mercedez, Ladder, and Non-planar (plus
their reduced forms obtained by shrinking some lines).  Now we have 10
\pagebreak[3]
topologies instead (figure~\ref{top}).  Each of them has 8
propagators.  In the diagrams with 4 heavy-quark lines
(figure~\ref{top}\emph{e}--\emph{g}), there is one linear dependence
between their denominators; with 5 heavy-quark lines
(figure~\ref{top}\emph{h}--\emph{j}) --- 2 dependences.  There are 9
independent scalar products of 3 loop momenta $k_{1,2,3}$ and the
4-velocity $v$.  Therefore, in the diagrams with two or three
heavy-quark lines (figure~\ref{top}\emph{a}--\emph{d}), there is one
scalar product in the numerator which cannot be cancelled against the
denominators; with 4 heavy-quark lines
\mbox{(figure~\ref{top}\emph{e}--\emph{g})} --- two scalar products;
with 5 heavy-quark lines (figure~\ref{top}\emph{h}--\emph{j}) ---
three scalar products.

When calculating these diagrams using recurrence relations,
some indices may vanish.
This corresponds to shrinking the corresponding lines.
In some cases, this results in diagrams with lower-loop propagator insertions,
which were calculated in section~\ref{Easy}.
The diagrams of figure~\ref{shrunk} are still non-trivial.

\subsection{Diagram with two heavy-quark lines}
\label{H3a}

Let's consider the diagram of figure~\ref{top}\emph{a} first.
We define
\begin{equation}
\begin{array}[b]{rclcrclcrcl}
\multicolumn{7}{r}{\displaystyle\int \frac{N^{n_0} d^d k_1 d^d k_2 d^d k_3}
{D_1^{n_1} D_2^{n_2} D_3^{n_3} D_4^{n_4} D_5^{n_5} D_6^{n_6} D_7^{n_7} D_8^{n_8}}=}
\nonumber\\[9pt]
&=&\multicolumn{9}{l}{\displaystyle-i \pi^{3d/2} (-2\omega)^{3d+2n_0-2\sum_{i=3}^8 n_i}
I_a(n_1,n_2,n_3,n_4,n_5,n_6,n_7,n_8;n_0)\,,}
\nonumber\\[5pt]
N   &=& \displaystyle \frac{k_3\cdot  v}{\omega}\,,&\qquad&
  D_1 &=& \displaystyle\frac{(k_1+p)\cdot v}{\omega}\,,&\qquad&
  D_2 &=& \displaystyle \frac{(k_2+p)\cdot  v}{\omega}\,,
\nonumber\\
D_3 &=& -k_1^2\,,&\qquad&
  D_4 &=& -k_2^2\,,&\qquad&
  D_5 &=& -(k_1-k_2)^2\,,
\nonumber\\
D_6 &=& -k_3^3\,,&\qquad&
  D_7 &=& -(k_3+k_1)^2\,,&\qquad&
  D_8 &=& -(k_3+k_2)^2\,.
\end{array}
\label{Ia}
\end{equation}
This integral is mirror-symmetric with respect to $1\leftrightarrow2$,
$3\leftrightarrow4$, $7\leftrightarrow8$.  It vanishes when the
indices of the following groups of lines are non-positive: 12, 67, 68,
78, 375, 485, 315, 425, 364, 137, 248, 157, or 258.
 
\FIGURE[t]{
\begin{picture}(132,114)
\put(66,59){\makebox(0,0){\epsfig{file=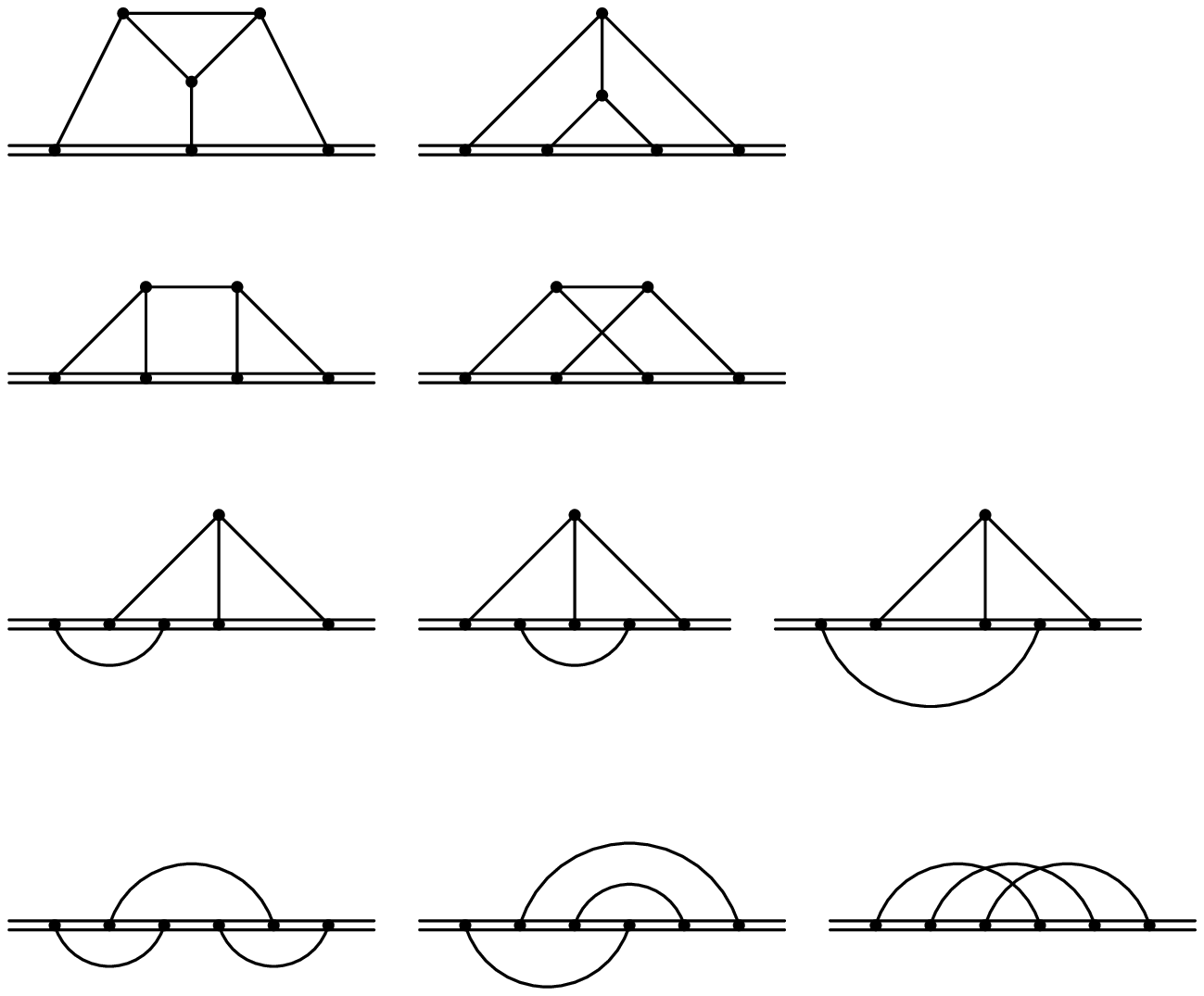}}}
\put(21,89){\makebox(0,0)[b]{$a$}}
\put(13.5,94.75){\makebox(0,0){1}}
\put(28.5,94.75){\makebox(0,0){2}}
\put(8,105){\makebox(0,0){3}}
\put(34,105){\makebox(0,0){4}}
\put(22.5,101){\makebox(0,0){5}}
\put(21,114.5){\makebox(0,0){6}}
\put(16,107){\makebox(0,0){7}}
\put(26,107){\makebox(0,0){8}}
\put(66,89){\makebox(0,0)[b]{$b$}}
\put(55.5,94.75){\makebox(0,0){1}}
\put(76.5,94.75){\makebox(0,0){2}}
\put(66,94.75){\makebox(0,0){3}}
\put(56,105){\makebox(0,0){4}}
\put(76,105){\makebox(0,0){5}}
\put(61,101){\makebox(0,0){6}}
\put(71,101){\makebox(0,0){7}}
\put(67.5,107){\makebox(0,0){8}}
\put(21,64){\makebox(0,0)[b]{$c$}}
\put(11,69.75){\makebox(0,0){1}}
\put(31,69.75){\makebox(0,0){2}}
\put(21,69.75){\makebox(0,0){3}}
\put(9,78){\makebox(0,0){4}}
\put(33,78){\makebox(0,0){5}}
\put(18,77){\makebox(0,0){6}}
\put(24,77){\makebox(0,0){7}}
\put(21,84.5){\makebox(0,0){8}}
\put(66,64){\makebox(0,0)[b]{$d$}}
\put(56,69.75){\makebox(0,0){1}}
\put(76,69.75){\makebox(0,0){2}}
\put(66,69.75){\makebox(0,0){3}}
\put(54,78){\makebox(0,0){4}}
\put(78,78){\makebox(0,0){5}}
\put(70,76){\makebox(0,0){6}}
\put(62,76){\makebox(0,0){7}}
\put(66,84.5){\makebox(0,0){8}}
\put(21,28){\makebox(0,0)[b]{$e$}}
\put(9,47.75){\makebox(0,0){1}}
\put(16,47.75){\makebox(0,0){2}}
\put(21,42.75){\makebox(0,0){3}}
\put(30,42.75){\makebox(0,0){4}}
\put(12,38.5){\makebox(0,0){5}}
\put(16,53){\makebox(0,0){6}}
\put(32,53){\makebox(0,0){7}}
\put(25.5,51){\makebox(0,0){8}}
\put(63,28){\makebox(0,0)[b]{$f$}}
\put(54,42.75){\makebox(0,0){1}}
\put(60,47.75){\makebox(0,0){2}}
\put(66,47.75){\makebox(0,0){3}}
\put(72,42.75){\makebox(0,0){4}}
\put(63,38.5){\makebox(0,0){5}}
\put(55,53){\makebox(0,0){6}}
\put(71,53){\makebox(0,0){7}}
\put(61,52){\makebox(0,0){8}}
\put(105,28){\makebox(0,0)[b]{$g$}}
\put(93,47.75){\makebox(0,0){1}}
\put(102,42.75){\makebox(0,0){2}}
\put(111,47.75){\makebox(0,0){3}}
\put(117,42.75){\makebox(0,0){4}}
\put(102,34){\makebox(0,0){5}}
\put(100,53){\makebox(0,0){6}}
\put(116,53){\makebox(0,0){7}}
\put(106,51){\makebox(0,0){8}}
\put(21,-2.75){\makebox(0,0)[b]{$h$}}
\put(9,14.75){\makebox(0,0){1}}
\put(16,14.75){\makebox(0,0){2}}
\put(21,9.75){\makebox(0,0){3}}
\put(26,14.75){\makebox(0,0){4}}
\put(33,14.75){\makebox(0,0){5}}
\put(12,5.5){\makebox(0,0){6}}
\put(30,5.5){\makebox(0,0){7}}
\put(21,21.25){\makebox(0,0){8}}
\put(66,-2.75){\makebox(0,0)[b]{$i$}}
\put(54,14.75){\makebox(0,0){1}}
\put(60,9.75){\makebox(0,0){2}}
\put(65,9.75){\makebox(0,0){3}}
\put(72,9.75){\makebox(0,0){4}}
\put(78,9.75){\makebox(0,0){5}}
\put(60,3.25){\makebox(0,0){6}}
\put(69,23.5){\makebox(0,0){7}}
\put(69,19){\makebox(0,0){8}}
\put(111,-2.75){\makebox(0,0)[b]{$j$}}
\put(99,9.75){\makebox(0,0){1}}
\put(105,9.75){\makebox(0,0){2}}
\put(111,9.75){\makebox(0,0){3}}
\put(117,9.75){\makebox(0,0){4}}
\put(123,9.75){\makebox(0,0){5}}
\put(104,21.25){\makebox(0,0){6}}
\put(111,21.25){\makebox(0,0){7}}
\put(118,21.25){\makebox(0,0){8}}
\end{picture}
\caption{Topologies of proper three-loop HQET propagator diagrams:
Mercedez (\emph{a, b, f, g, i}), Ladder (\emph{c, e, h}), Non-planar
(\emph{d, j})}
\label{top}}

\FIGURE[t]{
\begin{picture}(120,54)
\put(60,28){\makebox(0,0){\epsfig{file=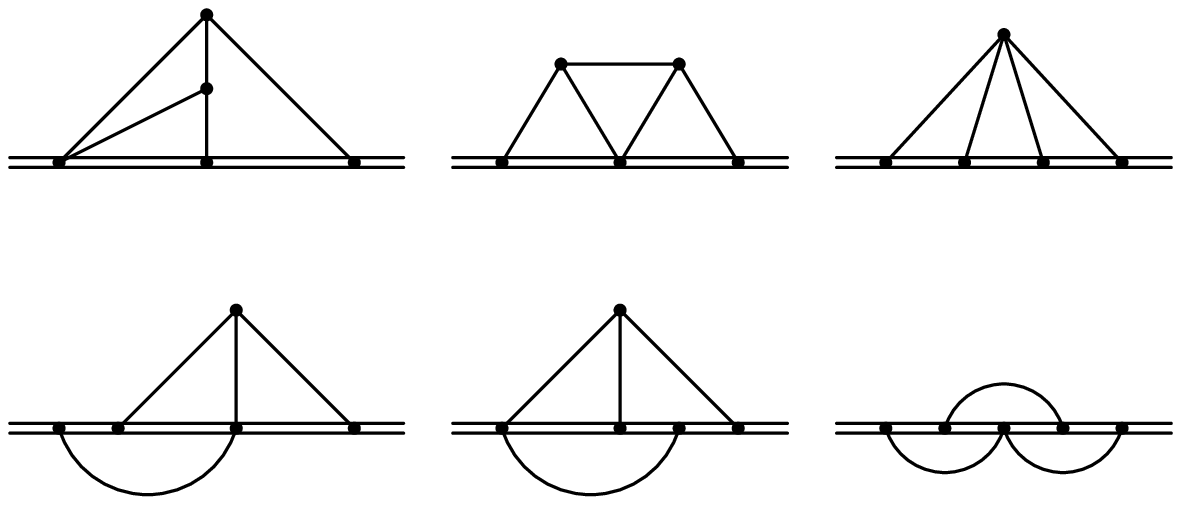}}}
\put(21,31.25){\makebox(0,0)[b]{$a$}}
\put(63,31.25){\makebox(0,0)[b]{$b$}}
\put(102,31.25){\makebox(0,0)[b]{$c$}}
\put(21,-2.5){\makebox(0,0)[b]{$d$}}
\put(63,-2.5){\makebox(0,0)[b]{$e$}}
\put(102,-2.5){\makebox(0,0)[b]{$f$}}
\end{picture}
\caption{Topologies of non-trivial three-loop diagrams with a shrunk line}
\label{shrunk}}

\clearpage

First we are going to get rid of the numerator.  When $n_0>0$ and
$n_7\ne1$, we can lower $n_0$ by
\begin{equation}
\left[ -2n_1\1+ + n_3\3+(\1--1) + n_5\5+(\1--\2-) + n_7\7+(\1--1+\0+)
\right] I_a = 0\,,
\label{a07}
\end{equation}
which is obtained by applying $\partial_1\cdot v$ to the integrand
of~(\ref{Ia}); the case $n_0>0$, $n_8\ne1$ is symmetric.  When $n_0>0$
and $n_6\ne1$, we can lower $n_0$ by
\begin{eqnarray}
\Bigl[2\left[3d+n_0-n_1-n_2-2(n_3+n_4+n_5+n_6+n_7+n_8)\right]
+\hphantom{\Bigr] I_a}&&
\nonumber\\
+ n_3\3+(\1--1) + n_4\4+(\2--1) - n_6\6+\0+ - 2n_0\0-
\Bigr] I_a &=& 0\,,
\label{a06}
\end{eqnarray}
which is the $(\partial_1+\partial_2-\partial_3)\cdot v$ relation
simplified using the homogeneity relation
\begin{equation}
\left[ 3d+n_0-n_1-n_2-2(n_3+n_4+n_5+n_6+n_7+n_8) + n_1\1+ + n_2\2+ \right] I_a = 0\,.
\label{ahom}
\end{equation}
When $n_0>0$ and $n_1\ne1$, we can lower $n_0$ by
\begin{eqnarray}
\Bigl[ d-n_1-n_3-n_5-2n_7 + n_1\1+(1-\0+)+\hphantom{\Bigr] I_a}&&
\nonumber\\
 + n_3\3+(\6--\7-) + n_5\5+(\8--\7-) \Bigr] I_a &=& 0\,,
\label{a01}
\end{eqnarray}
which is obtained by applying $\partial_1\cdot(k_1+k_3)$ to the
integrand of~(\ref{Ia}); the case $n_0>0$, $n_2\ne1$ is symmetric.  We
are left with $n_0>0$, $n_1=n_2=n_6=n_7=n_8=1$; we use
$\partial_3\cdot(k_3+k_1)$ relation
\begin{eqnarray}
\Bigl[ d+n_0-n_6-n_8-2n_7 + n_0\0-(\1--1)+\hphantom{\Bigr] I_a}&&
\nonumber\\
 + n_6\6+(\3--\7-) + n_8\8+(\5--\7-) \Bigr] I_a &=& 0
\label{at2}
\end{eqnarray}
to lower $n_0$ or raise $n_6$ or $n_8$, and apply the method described
above again.

Now we shall discuss the integral
$I_a(n_1,n_2,n_3,n_4,n_5,n_6,n_7,n_8)$ without numerator ($n_0=0$).
Applying $\partial_3\cdot k_3$, $\partial_1\cdot k_1$,
$\partial_1\cdot(k_1-k_2)$, $(\partial_1+\partial_2-\partial_3)\cdot
k_1$ to the integrand of~(\ref{Ia}), we obtain the recurrence
relations
\begin{eqnarray}
\left[ d-n_7-n_8-2n_6 + n_7\7+(\3--\6-) + n_8\8+(\4--\6-) \right] I_a &=& 0\,,
\label{at1}\\
\Bigl[ d-n_1-n_5-n_7-2n_3 + n_1\1+ +\hphantom{ \Bigr] I_a}&&
\nonumber\\
+\, n_5\5+(\4--\3-) + n_7\7+(\6--\3-) \Bigr] I_a &=& 0\,,
\label{at3}\\
\Bigl[ d-n_1-n_3-n_7-2n_5 + n_1\1+\2-+\hphantom{ \Bigr] I_a}&&
\nonumber\\
+\, n_3\3+(\4--\5-) + n_7\7+(\8--\5-) \Bigr] I_a &=& 0\,,
\label{at4}\\
\Bigl[ 2(d-n_5-n_7-n_8)-n_2-n_4-n_6 + n_2\2+\1- +\hphantom{ \Bigr] I_a}&&
\nonumber\\
+\, n_4\4+(\3--\5-) + n_6\6+(\3--\7-) \Bigr] I_a &=& 0\,,
\label{at5}
\end{eqnarray}
where the last relation was simplified using~(\ref{ahom}).
\pagebreak[3]

The cases $n_1=0$, $n_2=0$, $n_6=0$, $n_7=0$, $n_8=0$ are trivial.
When $n_1<0$, it can be raised by
\begin{eqnarray}
&&
\left[3d-n_1-n_2-2(n_3+n_4+n_5+n_6+n_7+n_8)\right] I_a =
\label{att}\\
&&\qquad
=\left[ d-n_1-n_2-n_4-n_8-2n_5 + n_4\4+(\3--\5-) + n_8\8+(\7--\5-) \right] \1+ I_a\,,
\nonumber
\end{eqnarray}
\looseness=1 which is the difference of~(\ref{ahom}) and $\1+$ shifted version of
the relation symmetric to~(\ref{at4}); the case $n_2<0$ is symmetric.
When $n_6<0$ and $n_7\ne1$, we can raise $n_6$ by~(\ref{at3}) (the
case $n_8\ne1$ is symmetric); when $n_6<0$ and $n_7=n_8=1$, we can
raise $n_6$ or $n_8$ by~(\ref{at2}).  When $n_7<0$ and $n_8\ne1$, we
can raise $n_7$ by the relation symmetric to~(\ref{at4}); when $n_7<0$
and $n_6\ne1$, we can raise $n_7$ by~(\ref{at5}); when $n_7<0$ and
$n_6=n_8=1$, we can raise $n_7$ or $n_6$ by the relation symmetric
to~(\ref{at2}).  The case $n_8<0$ is symmetric.  The case $n_3=0$
(figure~\ref{shrunk}\emph{a}, $J_a(n_1,n_2,n_4,n_5,n_6,n_7,n_8)$) will
be considered later in this section; the case $n_4=0$ is symmetric.
When $n_3<0$ and $n_6\ne1$, we can raise $n_3$ by~(\ref{at2}); when
$n_3<0$ and $n_7\ne1$, we can raise $n_3$ by~(\ref{at1}); when $n_3<0$
and $n_6=n_7=1$, we can raise $n_6$ or $n_7$ by the relation symmetric
to~(\ref{at2}).  The case $n_4<0$ is symmetric.  The case $n_5=0$
(figure~\ref{shrunk}\emph{b}, $J_b(n_1,n_2,n_3,n_4,n_6,n_7,n_8)$) will
be considered later in this section.  When $n_5<0$ and $n_8\ne1$, we
can raise $n_5$ by~(\ref{at2}) (the case $n_7\ne1$ is symmetric); when
$n_5<0$ and $n_7=n_8=1$, we can raise $n_7$ or $n_8$ by~(\ref{at1}).
When all the indices are positive, we can use~(\ref{at1}) to kill one
of the lines 3, 4, 6; or we can use~(\ref{at2}) to kill one of the
lines 3, 5, 7; or we can use the symmetric relation to kill one of the
lines 4, 5, 8; or we can use~(\ref{at5}) to kill one of the lines 1,
3, 5, 7; or we can use the symmetric relation to kill one of
the\linebreak lines 2, 4, 5, 8.

Now we consider $J_a(n_1,n_2,n_4,n_5,n_6,n_7,n_8)=
I_a(n_1,n_2,0,n_3,n_4,n_5,n_6,n_7,n_8)$
(figure~\ref{shrunk}\emph{a}µ).  This integral vanishes when the
indices of the following groups of lines are non-positive: 12, 67, 68,
78, 57, 15, 17, 46, 485, 425, 248, or 258.  If any index is zero, the
integral becomes trivial.  
If $n_1<0$, $n_2<0$, $n_6<0$, $n_7<0$, or $n_8<0$,
we just consider $J_a$ as $I_a$ with $n_3=0$ and proceed as usual;
the integral reduces to trivial ones not including $J_a$.
When $n_4>0$, we can kill one of the lines 2, 4, 8
using the relation symmetric to (\ref{at5}).
When $n_5>0$, we can kill one of the lines 2, 5, 8  using (\ref{at4}).
When both $n_4$ and $n_5$ are negative, we proceed as follows.
If $n_6\ne1$, $n_4$ can be raised by the relation symmetric to (\ref{at5});
if $n_7\ne1$, $n_5$ can be raised by (\ref{at4});
$n_2$ can be lowered down to 1 using (\ref{ahom}).
The remaining integral can be calculated by a repeated use of (\ref{I1m}):
\begin{eqnarray*}
\lefteqn{J_a(n_1,1,-|n_4|,-|n_5|,1,1,n_8) =}\qquad\qquad
\\
&=&\sum_{l_1,l_2,l_3,m} \frac{(-1)^{l_1+l_3} |n_4|! |n_5|!}
{(|n_5|-l_1)! m! (l_1-2m)! (|n_4|-l_2)! l_3! (l_2-l_3)!}\times
\\&&
\times\, I(n_1,1;l_1,m) I(1,n_8-|n_4|-|n_5|+l_1+l_2-m;l_2,0)\times
\\&&
\times\,I(-2(d+|n_4|+|n_5|-n_8)+n_1+l_3+3,1)\,.
\end{eqnarray*}

\looseness=1 Finally, we consider $J_b(n_1,n_2,n_3,n_4,n_6,n_7,n_8)=
I_a(n_1,n_2,n_3,n_4,0,n_6,n_7,n_8)$ (figure~\ref{shrunk}\emph{b}).
This integral is mirror-symmetric; it vanishes when the indices of the
following groups of lines are non-positive: 12, 67, 68, 78, 37, 48,
13, 17, 24, 28, or 364.  If any index is zero, the integral becomes
trivial.  If $n_1<0$, $n_2<0$, $n_3<0$, $n_4<0$, $n_6<0$, $n_7<0$, or
$n_8<0$, we just consider $J_b$ as $I_a$ with $n_5=0$ and proceed as
usual; the integral reduces to trivial ones not including $J_b$
(though, possibly, including $J_a$).  When all the indices are
positive, one of the lines 3, 4, 6 can be killed by~(\ref{at1}).

\subsection{Mercedez with three heavy-quark lines}
\label{H3b}

Now let's consider the diagram of figure~\ref{top}\emph{b}.
We define
\begin{equation}
\begin{array}[b]{rclcrclcrcl}
\multicolumn{5}{r}{\displaystyle\int \frac{N^{n_0} d^d k_1 d^d k_2 d^d k_3}
{D_1^{n_1} D_2^{n_2} D_3^{n_3} D_4^{n_4} D_5^{n_5} D_6^{n_6} D_7^{n_7} D_8^{n_8}}=}
\nonumber\\[9pt]
&=&\multicolumn{9}{l}{\displaystyle-i \pi^{3d/2} (-2\omega)^{3d+2n_0-2\sum_{i=4}^8 n_i}
I_b(n_1,n_2,n_3,n_4,n_5,n_6,n_7,n_8;n_0)\,,}
\nonumber\\[5pt]
\qquad D_1 &=& \displaystyle \frac{(k_1+p)\cdot  v}{\omega}\,,&\qquad&
  D_2 &=&\displaystyle \frac{(k_2+p)\cdot  v}{\omega}\,,&\qquad&
  D_3 &=&\displaystyle \frac{(k_3+p)\cdot v}{\omega}\,,
\nonumber\\
D_4 &=& -k_1^2\,,&\qquad&
  D_5 &=& -k_2^2\,,&\qquad&
  D_6 &=& -(k_1-k_3)^2\,,
\nonumber\\
D_7 &=& -(k_2-k_3)^2\,,&\qquad&
  D_8 &=& -(k_1-k_2)^2\,,&\qquad&
  N &=& -k_3^2\,.
\end{array}
\label{Ib}
\end{equation}
This integral is mirror-symmetric with respect to $1\leftrightarrow2$,
$4\leftrightarrow5$, $6\leftrightarrow7$.  It vanishes when the
indices of the following groups of lines are non-positive: 36, 37, 67, 123,
148, 168, 416, 486, 258, 278, 527, or 587.

First we are going to get rid of the numerator.  When $n_4\ne1$, we
can lower $n_0$ by the $\partial_1\cdot(k_1-k_3)$ relation
\begin{equation}
\left[ d-n_1-n_4-n_8-2n_6 + n_1\1+\3-
+ n_4\4+(\0+-\6-) + n_8\8+(\7--\6-) \right] I_b = 0\,;
\label{bt1}
\end{equation}
the case $n_5\ne0$ is symmetric.  When $n_6\ne1$, we can lower $n_0$
by the $\partial_1\cdot k_1$ relation
\begin{equation}
\left[ d-n_1-n_6-n_8-2n_4 + n_1\1+ + n_6\6+(\0+-\4-) + n_8\8+(\5--\4-)
\right] I_b = 0\,;
\label{bt2}
\end{equation}
the case $n_7\ne1$ is symmetric.  When $n_3\ne1$, we can lower $n_0$
or raise $n_6$ or $n_7$ by the $\partial_3\cdot v$ relation
\begin{equation}
\left[ -2n_3\3+ + n_0\0-(1-\3-) + n_6\6+(\3--\1-) + n_7\7+(\3--\2-)
\right] I_b = 0\,.
\label{bt3}
\end{equation}
Finally, we can lower $n_0$ or raise $n_3$ or $n_7$ by the
$\partial_3\cdot(k_3-k_1)$ relation
\begin{equation}
\left[ d+n_0-n_3-n_7-2n_6 + n_3\3+\1- + n_0\0-(\6--\4-) +
n_7\7+(\8--\6-) \right] I_b = 0\,.
\label{bt4}
\end{equation}

\pagebreak[3]

Now we shall discuss the integral
$I_b(n_1,n_2,n_3,n_4,n_5,n_6,n_7,n_8)$ without numerator ($n_0=0$).
Applying $\partial_1\cdot v$, $\partial_1\cdot(k_1-k_2)$,
$(\partial_1+\partial_2+\partial_3)\cdot k_1$ to the integrand
of~(\ref{Ib}), we obtain the recurrence relations
\begin{eqnarray}
\left[ -2n_1\1+ + n_4\4+(\1--1) + n_6\6+(\1--\3-) + n_8\8+(\1--\2-) \right] I_b &=& 0\,,
\label{bt5}\\[5pt]
\Bigl[ d-n_1-n_4-n_6-2n_8 + n_1\1+\2-+\hphantom{\Bigr] I_b}&&
\nonumber\\
+\, n_4\4+(\5--\8-) + n_6\6+(\7--\8-) \Bigr] I_b &=& 0\,,
\label{bt6}\\
\Bigl[ 2(d-n_6-n_7-n_8)-n_2-n_3-n_5 
+\hphantom{\Bigr] I_b}&&
\nonumber\\
+\, (n_2\2++n_3\3+)\1- + n_5\5+(\4--\8-) \Bigr] I_b &=& 0\,,
\label{bt7}
\end{eqnarray}
where the last relation was simplified using the homogeneity relation
\begin{eqnarray}
\Bigl[ 3d-n_1-n_2-n_3-2(n_4+n_5+n_6+n_7+n_8)+\hphantom{\Bigr] I_b}&&
\nonumber\\
 +\, n_1\1+ + n_2\2+ + n_3\3+ \Bigr] I_b &=& 0\,.
\label{bhom}
\end{eqnarray}
Using it to simplify the sum of~(\ref{bt5}), its symmetric and~(\ref{bt3}), we get
\begin{eqnarray}
\Bigl[ 2\left[3d-n_1-n_2-n_3-2(n_4+n_5+n_6+n_7+n_8)\right]
+\hphantom{\Bigr] I_b}&&
\nonumber\\
+\, n_4\4+(\1--1) + n_5\5+(\2--1) \Bigr] I_b &=& 0\,.
\label{bt8}
\end{eqnarray}

\looseness=1 The cases $n_3=0$, $n_4=0$, $n_5=0$, $n_6=0$, $n_7=0$ are trivial.
When $n_3<0$ and $n_6\ne1$, we can raise $n_3$ by~(\ref{bt5})
(the case $n_7\ne1$ is symmetric);
when $n_3<0$ and $n_6=n_7=1$, we can raise $n_3$ or $n_7$ by~(\ref{bt4}).
When $n_4<0$ and $n_5\ne1$, we can raise $n_4$
by the relation symmetric to~(\ref{bt6});
when $n_4<0$ and $n_5=1$, we can raise $n_4$ or $n_5$ by~(\ref{bt8}).
The case $n_5<0$ is symmetric.
When $n_6<0$ and $n_7\ne1$, we can raise $n_6$ by~(\ref{bt4});
when $n_6<0$ and $n_3\ne1$, $n_7=1$,
we can raise $n_6$ or $n_7$ by~(\ref{bt3});
when $n_6<0$ and $n_3=n_7=1$, we can raise $n_6$ or $n_3$
by the relation symmetric to~(\ref{bt4}).
The case $n_7<0$ is symmetric.
The case $n_1=0$ is $J_a(n_3,n_2,n_5,n_7,n_4,n_6,n_8)$
(figure~\ref{shrunk}\emph{a}, section~\ref{H3a}), $n_2=0$ is symmetric.
When $n_1<0$ and $n_8\ne1$,
we can raise $n_1$ by the relation symmetric to~(\ref{bt5});
when $n_1<0$ and $n_6\ne1$, we can raise $n_1$ by~(\ref{bt3});
when $n_1<0$ and $n_4\ne1$, we can raise $n_1$ by~(\ref{bt8});
when $n_1<0$ and $n_2\ne1$,
we can raise $n_1$ by the relation symmetric to~(\ref{bt6});
when $n_1<0$ and $n_3\ne1$, we can raise $n_1$ by~(\ref{bt4});
when $n_1<0$ and $n_2=n_3=n_4=n_6=n_8=1$,
we can raise $n_1$, $n_2$ or $n_3$ by~(\ref{bhom}).
The case $n_2<0$ is symmetric.
The case $n_8=0$ (figure~\ref{shrunk}\emph{c}, $J_c(n_1,n_2,n_3,n_4,n_5,n_6,n_7)$)
will be considered later in this section.
When $n_8<0$ and $n_7\ne1$, we can raise $n_8$ by~(\ref{bt4})
(the case $n_6\ne1$ is symmetric);
\pagebreak[3]
when $n_8<0$ and $n_5\ne1$, we can raise $n_8$ by~(\ref{bt7})
(the case $n_4\ne1$ is symmetric);
when $n_8<0$ and $n_1\ne1$, we can raise $n_8$, $n_4$ or $n_6$ by~(\ref{bt5})
(the case $n_2\ne1$ is symmetric);
when $n_8<0$ and $n_3\ne1$, we can raise $n_6$ or $n_7$ by~(\ref{bt3});
when $n_8<0$ and $n_1=n_2=n_3=n_4=n_5=n_6=n_7=1$,
we can raise $n_1$, $n_2$ or $n_3$ by~(\ref{bhom}).
When all the indices are positive,
we can kill one of the lines 1, 6, 8 by~(\ref{bt4}),
or one of the lines 2, 7, 8 by its mirror-symmetric relation.

Now we consider $J_c(n_1,n_2,n_3,n_4,n_5,n_6,n_7)=
I_b(n_1,n_2,n_3,n_4,n_5,n_6,n_7,n_8)$ (figure~\ref{shrunk}\emph{c}).
It is mirror-symmetric;
it vanishes if any two indices of $n_4$, $n_5$, $n_6$, $n_7$ are non-positive,
or $n_1$, $n_2$, $n_3$ are all non-positive.
If any index is zero, the integral becomes trivial.
If any index is negative, we just consider $J_c$ as $I_b$ with $n_8=0$
and proceed as usual.
Using the $\partial_1\cdot v$ and $\partial_3\cdot v$ relations
\begin{eqnarray}
\left[ - 2n_1\1+ + n_4\4+(\1--1) + n_6\6+(\1--\3-) \right] J_c &=& 0\,,
\label{jc1}\\[3pt]
\left[ - 2n_3\3+ + n_6\6+(\3--\1-) + n_7\7+(\3--\2-) \right] J_c &=& 0\,,
\label{jc2}
\end{eqnarray}
we can lower $n_1$, $n_2$ (symmetric case) and $n_3$ down to 1.

We are left with $J_c(1,1,1,n_4,n_5,n_6,n_7)$.  It is rather difficult
to apply the standard techniques to this integral, because each
operator $\partial_i\cdot k_j$ produces a scalar product which cannot
be expressed via the denominators.  For example, $\partial_1\cdot k_1$
gives the term $-\frac{n_6}{D_6}2k_1\cdot k_3$, and
$\partial_1\cdot(k_1-k_3)$ gives the term $-\frac{n_4}{D_4}2k_1\cdot
k_3$.  We can cancel $2k_1\cdot k_3$ by forming the difference
\[
(n_4-1) \6- \partial_1\cdot k_1 - (n_6-1) \4- \partial_1\cdot(k_1-k_3)\,,
\]
which results in
\begin{equation}
\Bigl[ (n_4-1) \left( d-n_1-2n_4 + n_1\1+ \right) \6- 
-\, (n_6-1) \left( d-n_1-2n_6 + n_1\1+\3- \right) \4- \Bigr] J_c = 0\,.
\end{equation}
Another useful combination is
\[
(n_4-1) \5- \left(\partial_1+\partial_2+\partial_3\right)\cdot k_1
- (n_5-1) \4- \left(\partial_1+\partial_2+\partial_3\right)\cdot k_2
\]
(it can be simplified by the homogeneity relation).
Using~(\ref{jc1}) and~(\ref{jc2}), we obtain at $n_1=n_2=n_3=1$
\begin{eqnarray}
\Bigl[ (n_4-1) \left[2(d-2n_4-1)-n_4\4+(1-\1-)\right]\6- 
-\hphantom{\Bigr] J_c}&&
\nonumber\\[2pt]
{}- 2(n_6-1)(d-2n_6-1+\1+\3-)\4-
+ (n_4-1)(n_6-1)(\1--\3-) \Bigr] J_c &=& 0\,,\qquad
\label{jc4}\\[5pt]
\Bigl[ (n_4-1)\left[2(d-n_5-n_6-n_7)+(\2++\3+)\1-\right]\5-
-\hphantom{\Bigr] J_c}&&
\nonumber\\[2pt]
{}- (n_5-1) \left[2(d-n_4-n_6-n_7)+(\1++\3+)\2-\right]\4- \Bigr] J_c &=& 0\,.\qquad
\label{jc6}
\end{eqnarray}
One more useful relation is obtained by adding~(\ref{jc1}), its mirror-symmetric,
formula~(\ref{jc2}) and using the homogeneity relation:
\begin{eqnarray}
\Bigl[ 2\left[3d-n_1-n_2-n_3-2(n_4+n_5+n_6+n_7)\right]+
\hphantom{\Bigr] J_c}&&
\nonumber\\[2pt]
 +\, n_4\4+(\1--1) + n_5\5+(\2--1) \Bigr] J_c &=& 0\,.
\label{jc3}
\end{eqnarray}
When $n_6>1$, we can lower it down to 1 (raising $n_4$) by~(\ref{jc4});
the case $n_7>1$ is symmetric.
Finally, we can lower $n_4$ and $n_5$ down to 1
by~(\ref{jc6}) together with~(\ref{jc3}).

The integral $J_c(1,1,1,1,1,1,1)$ cannot be reduced to simpler ones
(in contrast to the massless case~\cite{CT}), and should be considered
as a new basis integral.  Its value is currently unknown, and its
calculation is highly non-trivial.

\subsection{Ladder with three heavy-quark lines}
\label{H3c}

Now let's consider the diagram of figure~\ref{top}\emph{c}.
We define
\begin{equation}
\begin{array}[b]{rclcrclcrcl}
\multicolumn{5}{r}{\displaystyle\int \frac{N^{n_0} d^d k_1 d^d k_2 d^d k_3}
{D_1^{n_1} D_2^{n_2} D_3^{n_3} D_4^{n_4} D_5^{n_5} D_6^{n_6} D_7^{n_7} D_8^{n_8}}=}
\nonumber\\[9pt]
&=&\multicolumn{9}{l}{ \displaystyle-i \pi^{3d/2} (-2\omega)^{3d+2n_0-2\sum_{i=4}^8 n_i}
I_c(n_1,n_2,n_3,n_4,n_5,n_6,n_7,n_8;n_0)\,,}
\nonumber\\[5pt]
\qquad\qquad D_1 &=& \displaystyle \frac{(k_1+p)\cdot  v}{\omega}\,,&\qquad&
  D_2 &=&\displaystyle \frac{(k_2+p)\cdot  v}{\omega}\,,&\qquad&
  D_3 &=&\displaystyle \frac{(k_3+p)\cdot v}{\omega}\,,
\nonumber\\
D_4 &=& -k_1^2\,,&\qquad&
  D_5 &=& -k_2^2\,,&\qquad&
  D_6 &=& -(k_1-k_3)^2\,,
\nonumber\\
D_7 &=& -(k_2-k_3)^2\,,&\qquad&
  D_8 &=& -k_3^2\,,&\qquad&
  N &=& 2k_1\cdot k_2\,.
\end{array}
\label{Ic}
\end{equation}
This integral is mirror-symmetric with respect to $1\leftrightarrow2$,
$4\leftrightarrow5$, $6\leftrightarrow7$.  It vanishes when the
indices of the following groups of lines are non-positive: 14, 16, 46,
25, 27, 57, 132, 637, 368, 378, 687, 485, 487, 586.

First we are going to get rid of the numerator.  When $n_7\ne1$, we
can lower $n_0$ by the $\partial_3\cdot(k_3-k_1)$ relation
\begin{eqnarray}
\Bigl[ d-n_3-n_7-n_8-2n_6 + n_3\3+\1-+\hphantom{\Bigr] I_c}&&
\nonumber\\
+\, n_7\7+(\0++\4-+\5--\6-) + n_8\8+(\4--\6-) \Bigr]
I_c &=& 0\,;
\label{ct1}
\end{eqnarray}
the case $n_6\ne0$ is symmetric.  When $n_5\ne1$, we can lower $n_0$ by
\begin{eqnarray}
\Bigl[ 2(d-n_5-n_6-n_7)+n_0-n_2-n_3-n_8 + 2n_0\0-\4- + 
\hphantom{\Bigr] I_c}&&
\nonumber\\
+\,(n_2\2++n_3\3+)\1- - n_5\5+\0+ + n_8\8+(\4--\6-) \Bigr] I_c &=& 0\,,
\label{ct2}
\end{eqnarray}
which is the $(\partial_1+\partial_2+\partial_3)\cdot k_1$ relation
simplified by the homogeneity relation.  the case $n_4\ne1$ is
symmetric.  When $n_1\ne1$, we can lower $n_0$ or raise $n_4$ or $n_6$
by the $\partial_1\cdot v$ relation
\begin{equation}
\left[ -2n_1\1+ + n_0\0-(\2--1) + n_4\4+(\1--1) + n_6\6+(\1--\3-)
\right] I_c = 0\,;
\label{ct3}
\end{equation}
the case $n_2\ne1$ is symmetric.  Finally, we can raise $n_1$ or $n_6$
by the $\partial_1\cdot k_1$ relation
\begin{equation}
\left[ d+n_0-n_1-n_6-2n_4 + n_1\1+ + n_6\6+(\8--\4-) \right] I_c = 0\,.
\label{ct4}
\end{equation}

Now we shall discuss the integral
$I_c(n_1,n_2,n_3,n_4,n_5,n_6,n_7,n_8)$ without numerator ($n_0=0$).
Applying $\partial_3\cdot v$, $\partial_1\cdot(k_1-k_3)$ to the
integrand of~(\ref{Ic}), we obtain the recurrence relations
\begin{eqnarray}
\left[ -2n_3\3+ + n_6\6+(\3--\1-) + n_7\7+(\3--\2-) + n_8\8+(\3--1) \right] I_c &=& 0\,,
\label{ct5}\\
\left[ d-n_1-n_4-2n_6 + n_1\1+\3- + n_4\4+(\8--\6-) \right] I_c &=& 0\,.\qquad
\label{ct6}
\end{eqnarray}
Homogeneity in $\omega$ gives the relation identical with~(\ref{bhom}).
\pagebreak[3]

The cases $n_1=0$, $n_2=0$, $n_4=0$, $n_5=0$, $n_6=0$, $n_7=0$ are trivial.
When $n_1<0$ and $n_6\ne1$, we can raise $n_1$ by~(\ref{ct5});
when $n_1<0$ and $n_4\ne1$, we can raise $n_1$
using the sum of~(\ref{ct3}) and~(\ref{ct5});
when $n_1<0$ and $n_4=n_6=1$,
we can raise $n_1$ or $n_6$ by~(\ref{ct4}).
The case $n_2<0$ is symmetric.
When $n_4<0$ and $n_6\ne1$, we can raise $n_4$ by~(\ref{ct4});
when $n_4<0$ and $n_1\ne1$, $n_6=1$,
we can raise $n_4$ or $n_6$ by~(\ref{ct3});
when $n_4<0$ and $n_1=n_6=1$,
we can raise $n_4$ or $n_1$ by~(\ref{ct6}).
The case $n_5<0$ is symmetric.
When $n_6<0$ and $n_4\ne1$, we can raise $n_6$ by~(\ref{ct6});
when $n_6<0$ and $n_1\ne1$, $n_4=1$,
we can raise $n_6$ or $n_4$ by~(\ref{ct3});
when $n_6<0$ and $n_1=n_4=1$, we can raise $n_6$ or $n_1$ by~(\ref{ct4}).
The case $n_7<0$ is symmetric.
The case $n_3=0$ is $J_b(n_1,n_2,n_4,n_5,n_8,n_6,n_7)$
(figure~\ref{shrunk}\emph{a}, section~\ref{H3a}).
When $n_3<0$ and $n_6\ne1$, we can raise $n_3$ by~(\ref{ct3})
(the case $n_7\ne1$ is symmetric);
when $n_3<0$ and $n_8\ne1$, we can raise $n_3$ by the relation
\begin{eqnarray}
\Bigl[ 2\left[3d-n_1-n_2-n_3-2(n_4+n_5+n_6+n_7+n_8)\right]+
\hphantom{\Bigr] I_c}&&
\nonumber\\
 +\, n_4\4+(\1--1) + n_5\5+(\2--1) + n_8\8+(\3--1) \Bigr] I_c &=&0\,,
\label{csum}
\end{eqnarray}
which is~(\ref{ct3}) plus its mirror-symmetric plus~(\ref{ct5})
simplified by the homogeneity relation;
when $n_3<0$ and $n_1\ne1$, we can raise $n_3$ by~(\ref{ct6})
(the case $n_2\ne1$ is symmetric);
when $n_3<0$ and $n_1=n_2=n_6=n_8=1$,
we can raise $n_1$, $n_2$ or $n_3$ by the homogeneity relation.
The case $n_8=0$ is $J_c(n_1,n_2,n_3,n_4,n_5,n_6,n_7)$
(figure~\ref{shrunk}\emph{c}, section~\ref{H3b}).
When $n_8<0$ and $n_6\ne1$, we can raise $n_8$ by~(\ref{ct4})
(the case $n_7\ne1$ is symmetric);
when $n_8<0$ and $n_4\ne1$, we can raise $n_8$ by~(\ref{ct6})
(the case $n_5\ne1$ is symmetric);
when $n_8<0$ and $n_3\ne1$, $n_4=n_6=n_7=1$,
we can raise $n_8$, $n_6$ or $n_7$ by~(\ref{ct5});
when $n_8<0$ and $n_1\ne1$, $n_3=n_4=n_5=n_6=n_7=1$,
we can raise $n_4$ or $n_6$ by~(\ref{ct3})
(the case $n_2\ne1$ is symmetric);
when $n_8<0$ and $n_1=n_2=n_3=n_4=n_5=n_6=n_7=1$,
we can raise $n_1$, $n_2$ or $n_3$ the homogeneity relation.
When all the indices are positive,
we can kill one of the lines 3, 6, 8 by~(\ref{ct6}),
or one of the lines 3, 7, 8 by its mirror-symmetric relation.

\subsection{Non-planar diagram with three heavy-quark lines}
\label{H3d}

Now let's consider the diagram of figure~\ref{top}\emph{d}.
We define
\begin{equation}
\begin{array}[b]{rclcrclcrcl}
\multicolumn{7}{r}{\displaystyle\int \frac{N^{n_0} d^d k_1 d^d k_2 d^d k_3}
{D_1^{n_1} D_2^{n_2} D_3^{n_3} D_4^{n_4} D_5^{n_5} D_6^{n_6} D_7^{n_7} D_8^{n_8}}=\qquad\qquad}
\nonumber\\[9pt]
&=&\multicolumn{9}{l}{ \displaystyle-i \pi^{3d/2} (-2\omega)^{3d+2n_0-2\sum_{i=4}^8 n_i}
I_d(n_1,n_2,n_3,n_4,n_5,n_6,n_7,n_8;n_0)\,,}
\nonumber\\[5pt]
D_1 &=&\displaystyle \frac{(k_1+p)\cdot v}{\omega}\,,&\qquad&
  D_2 &=&\displaystyle  \frac{(k_2+p)\cdot v}{\omega}\,,&\qquad&
D_3 &=&\displaystyle  \frac{(k_1+k_2-k_3+p)\cdot v}{\omega}\,,
\nonumber\\
  D_4 &=& -k_1^2\,,&\qquad&
  D_5 &=& -k_2^2\,,&\qquad&
D_6 &=& -(k_1-k_3)^2\,,
\label{Id}\\
  D_7 &=& -(k_2-k_3)^2\,,&\qquad&
  D_8 &=& -k_3^2\,,&\qquad&
  N &=& 2k_1\cdot k_2\,.
\nonumber
\end{array}
\end{equation}
This integral is mirror-symmetric with respect to $1\leftrightarrow2$,
$4\leftrightarrow5$, $6\leftrightarrow7$.  It vanishes when the
indices of the following groups of lines are non-positive: 46, 57,
132, 485, 487, 586, 687, 314, 136, 325, 237, 148, 178, 258, 268, 417,
526, 637, 368, 378.

When $n_1\le0$, $n_2\le0$ or $n_3\le0$, we use
\begin{eqnarray}
&&I_d(-|n_1|,n_2,n_3,n_4,n_5,n_6,n_7,n_8;n_0) =
\nonumber\\
&&= (\0++\1-)^{|n_1|} (\5-+\6--\3--\8-)^{n_0}
I_a(n_3,n_2,0,n_5,n_6,n_7,n_4,n_8)\,,\qquad
\label{id2ia1}\\
&&I_d(n_1,n_2,-|n_3|,n_4,n_5,n_6,n_7,n_8;n_0) =
\nonumber\\
&&= (\0++\1-+\2--1)^{|n_3|} (\5--\3--\4-)^{n_0}
I_a(n_1,n_2,n_4,n_5,0,n_8,n_6,n_7)\,,\qquad
\label{id2ia2}
\end{eqnarray}
and relation symmetric to~(\ref{id2ia1}).
Applying $\partial_2\cdot v$ and $-\partial_3\cdot v$
to the integrand of~(\ref{Id}), we obtain
\begin{eqnarray}
\Bigl[ 2 n_2\2+ + 2 n_3\3+ + n_5\5+(1-\2-) + n_7\7+(\1--\3-) \Bigr] I_d = 0\,,
\label{dt1}\\[5pt]
\Bigl[ 2n_3\3+ + n_6\6+(\2--\3-) + n_7\7+(\1--\3-)
+\hphantom{\Bigr] I_d}&&
\nonumber\\
+\, n_8\8+(\1-+\2--\3--1) \Bigr] I_d &=& 0\,.
\label{dt2}
\end{eqnarray}
When $n_1>1$, we can lower it by
\begin{eqnarray}
\Bigl[2\bigl[3d-n_1-n_2-n_3-2(n_4+n_5+n_6+n_7+n_8-n_0)\bigr]
+ \hphantom{\Bigr] I_d}
\nonumber\\
{} + n_0\0-(\1--1) + 2 n_1\1+ + n_5\5+(\2--1) + n_7\7+(\3--\1-) \Bigr] I_d &=& 0\,,
\label{dt8}
\end{eqnarray}
which is twice the homogeneity relation
\begin{eqnarray}
\Bigl[3d-n_1-n_2-n_3-2(n_4+n_5+n_6+n_7+n_8-n_0)+\hphantom{\Bigr] I_d}&&
\nonumber\\
 +\, n_1\1+ + n_2\2+ + n_3\3+ \Bigr] I_d &=& 0
\label{dhom}
\end{eqnarray}
minus~(\ref{dt1}); the case $n_2>1$ is symmetric.
When $n_3>1$, we can lower it by~(\ref{dt2}).

We are left with $I_d(1,1,1,n_4,n_5,n_6,n_7,n_8;n_0)$,
and now are going to get rid of the numerator.
Applying $\partial_3\cdot(k_3-k_2)$ and $\partial_1\cdot(k_1-k_3)$
to the integrand of~(\ref{Id}), we obtain
\begin{eqnarray}
\Bigl[ d-n_3-n_6-n_8-2n_7 + n_3\3+\1- + \hphantom{\Bigr]I_d}&&
\nonumber\\
{} + n_6\6+(\4-+\5--\7-+\0+) + n_8\8+(\5--\7-) \Bigr] I_d &=& 0\,,
\label{dt3}\\[5pt]
\Bigl[ d+n_0-n_3-n_4-2n_6 + n_0\0-(\5--\7-+\8-) + \hphantom{\Bigr]I_d}&&
\nonumber\\
{} + n_1\1+(\2--\3-) + n_3\3+\2- + n_4\4+(\8--\6-) \Bigr] I_d &=& 0\,.
\label{dt4}
\end{eqnarray}
When $n_4\ne1$, we can lower $n_0$ by
\begin{eqnarray}
\Bigl[2(d-n_4-n_6-n_7)+n_0-n_1-n_2-n_3-n_8+1 - \hphantom{\Bigr]I_d}&&
\nonumber\\
{} - \bigl[3d-n_1-n_2-n_3+1-2(n_4+n_5+n_6+n_7+n_8-n_0)\bigr]\2-
+ \hphantom{\Bigr]I_d}&&
\nonumber\\
{} + 2 n_0\0-\5- - n_4\4+\0+ + n_8\8+(\5--\7-) \Bigr] I_d &=& 0\,,\qquad
\label{dt9}
\end{eqnarray}
which is the $-(\partial_1+\partial_2+\partial_3)\cdot k_2$ relation
plus~(\ref{dhom}) minus $\2-$ shifted~(\ref{dhom});
the case $n_5\ne1$ is symmetric.
When $n_6\ne1$, we can lower $n_0$ by~(\ref{dt3}); the case $n_7\ne1$ is symmetric.
Finally, when $n_4=n_5=n_6=n_7=1$, we can lower $n_0$ or raise $n_4$ by~(\ref{dt4}).

Now we shall discuss the integral $I_d(n_1,n_2,n_3,n_4,n_5,n_6,n_7,n_8)$
without numerator ($n_0=0$).
Applying $\partial_1\cdot k_1$, $\partial_3\cdot k_3$, $(\partial_2+\partial_3)\cdot k_3$
to the integrand of~(\ref{Id}), we obtain
\begin{eqnarray}
\Bigl[ d-n_1-n_6-2n_4 + n_1\1+ + n_3\3+(1-\1-) + n_6\6+(\8--\4-) \Bigr] I_d &=& 0\,,
\label{dt5}\\[5pt]
\Bigl[ d-n_3-n_6-n_7-2n_8 + n_3\3+(\1-+\2--1) + \hphantom{\Bigr] I_d}&&
\nonumber\\
{} + n_6\6+(\4--\8-) + n_7\7+(\5--\8-) \Bigr] I_d &=& 0\,,
\label{dt6}\\[5pt]
\Bigl[ d-n_2-n_5-n_6-2n_8 + n_2\2+(1+\3--\1-) + \hphantom{\Bigr] I_d}&&
\nonumber\\
{} + n_5\5+(\7--\8-) + n_6\6+(\4--\8-) \Bigr] I_d &=& 0\,.
\label{dt7}
\end{eqnarray}
Some other useful relations are
\begin{equation}
\Bigl[2 n_2\2+ + n_5\5+(1-\2-) + n_6\6+(\3--\2-) + n_8\8+(1+\3--\1--\2-)\Bigr] I_d = 0\,,
\label{dt14}
\end{equation}
\begin{equation}
\Bigl[ 2(d-n_5-n_7-n_8)-n_2-n_3-n_6 + n_2\2+ + n_3\3+\1- + n_6\6+(\4--\8-) \Bigr] I_d = 0\,,
\label{dt13}
\end{equation}
the differences of~(\ref{dt1}) and~(\ref{dt2}), and of~(\ref{dhom}) and~(\ref{dt5}).

The case $n_1=0$ is $J_a(n_3,n_2,n_5,n_6,n_7,n_4,n_8)$
(figure~\ref{shrunk}\emph{a}, section~\ref{H3a}; $n_2=0$ is symmetric);
the case $n_3=0$ is $J_b(n_1,n_2,n_4,n_5,n_8,n_6,n_7)$
(figure~\ref{shrunk}\emph{b}, section~\ref{H3a});
the case $n_8=0$ is $J_c(n_1,n_2,n_3,n_4,n_5,n_7,n_6)$
(figure~\ref{shrunk}\emph{c}, section~\ref{H3b}).
The cases $n_6=0$
($J_d(n_1,n_3,n_2,n_4,n_7,n_5,n_8)$, figure~\ref{shrunk}\emph{d}; $n_7=0$ is symmetric)
and $n_4=0$
($J_e(n_1,n_3,n_2,n_6,n_8,n_5,n_7)$, figure~\ref{shrunk}\emph{e}; $n_5=0$ is symmetric)
will be discussed later in this section.
When $n_2<0$, it can be raised using
\begin{eqnarray}
&&\bigl[3d-n_1-n_2-n_3-2(n_4+n_5+n_6+n_7+n_8)\bigr] I_d =
\nonumber\\
&&= \bigl[d-n_2-n_3-n_4-2n_6 - n_1\1+\3- + n_4\4+(\8--\6-)\bigr] \2+ I_d\,,
\label{dt10}
\end{eqnarray}
which is~(\ref{dhom}) minus $\2+$ shifted~(\ref{dt4});
the case $n_1<0$ is symmetric.
When $n_3<0$ and $n_1\ne1$, we can raise $n_3$ by~(\ref{dt4}) ($n_2\ne1$ is symmetric);
when $n_3<0$ and $n_7\ne1$, we can raise $n_3$ by~(\ref{dt1}) ($n_6\ne1$ is symmetric);
when $n_3<0$ and $n_8\ne1$, we can raise $n_3$ using
\begin{eqnarray}
\Bigl[ 2\bigl[3d-n_1-n_2-n_3-2(n_4+n_5+n_6+n_7+n_8)\bigl] + \hphantom{\Bigr] I_d}&&
\nonumber\\
{} + n_4\4+(\1--1) + n_5\5+(\2--1) + n_8\8+(\1-+\2--\3--1) \Bigr] I_d &=& 0\,,
\label{dt11}
\end{eqnarray}
which is~(\ref{dt1}) plus its symmetric minus~(\ref{dt2}) simplified by~(\ref{dhom});
when $n_3<0$ and $n_1=n_2=n_6=n_7=n_8=1$,
we can raise $n_3$, $n_1$ or $n_2$ by~(\ref{dhom}).
When $n_1>1$, it can be lowered by~(\ref{dt8}); the case $n_2>1$ is symmetric.
When $n_3>1$, it can be lowered by~(\ref{dt2}).

We are left with $I_d(1,1,1,n_4,n_5,n_6,n_7,n_8)$.
When $n_4<0$, it can be raised by~(\ref{dt4}); the case $n_5<0$ is symmetric.
When $n_6<0$ and $n_4\ne1$, we can raise $n_6$ by~(\ref{dt4});
when $n_6<0$ and $n_4=1$, we can raise $n_6$ or $n_4$ using
\begin{eqnarray}
\Bigl[ 2(d-n_1-n_6-2n_4) - 2 n_3\3+\1- + n_4\4+(\1--1) + \hphantom{\Bigr] I_d}&&
\nonumber\\
{} + n_6\6+ \bigl[2(\8--\4-)+\3--\2-\bigr] \Bigr] I_d &=& 0\,,
\label{dt12}
\end{eqnarray}
which is twice~(\ref{dt5}) minus the relation symmetric to~(\ref{dt1}).
The case $n_7<0$ is symmetric.
When $n_8<0$ and $n_4\ne1$, we can raise $n_8$ by~(\ref{dt4}) ($n_5\ne1$ is symmetric);
when $n_8<0$ and $n_6\ne1$, we can raise $n_8$ by~(\ref{dt12}) ($n_7\ne1$ is symmetric);
when $n_8<0$ and $n_1=n_2=n_3=n_4=n_5=n_6=n_7=1$,
we can raise $n_8$, $n_4$ or $n_5$ by~(\ref{dt11}).
When all the indices are positive,
we can kill the line 6 or 8 using~(\ref{dt4}).

Thus, the non-planar diagram reduces to planar ones, in contrast to
the massless case where such a reduction is impossible~\cite{CT,Ba}.

Now we consider $J_d(n_1,n_2,n_3,n_4,n_5,n_7,n_8)=
I_d(n_1,n_2,n_3,n_4,n_5,0,n_7,n_8)$ (figure~\ref{shrunk}\emph{d}).
This integral vanishes when indices of the following groups of lines
are non-positive: 4, 13, 23, 57, 58, 78, 37, 38, 25, 28.
It becomes trivial if any of the indices is zero.
Applying $(\partial_2+\partial_3)\cdot k_2$ and $\partial_3\cdot(k_3-k_2)$
to the integrand of~(\ref{Id}), we obtain
\begin{eqnarray}
\Bigl[ d-n_2-n_8-2n_5 + n_2\2+ + n_8\8+(\7--\5-) \Bigr] J_d &=& 0\,,
\label{dt15}\\[2pt]
\Bigl[ d-n_3-n_8-2n_7 + n_3\3+\1- + n_8\8+(\5--\7-) \Bigr] J_d &=& 0\,.
\label{dt16}
\end{eqnarray}
When $n_3<0$ and $n_2\ne1$, we can raise $n_3$ by~(\ref{dt7});
when $n_3<0$ and $n_7\ne1$, we can raise $n_3$ by~(\ref{dt8});
when $n_3<0$ and $n_8\ne1$, we can raise $n_3$ by~(\ref{dt14});
when $n_3<0$ and $n_2=n_7=n_8=1$, we can raise $n_3$ or $n_2$ by~(\ref{dt13}).
When $n_2<0$ and $n_3\ne1$, we can raise $n_2$ by~(\ref{dt6});
when $n_2<0$ and $n_5\ne1$, we can raise $n_2$ by~(\ref{dt8});
when $n_2<0$ and $n_8\ne1$, we can raise $n_2$ by~(\ref{dt2});
when $n_2<0$ and $n_3=n_5=n_8=1$, we can raise $n_2$ or $n_3$ by~(\ref{dt13}).
When $n_8<0$ and $n_7\ne1$, we can raise $n_8$ by~(\ref{dt6});
when $n_8<0$ and $n_5\ne1$, we can raise $n_8$ by~(\ref{dt7});
when $n_8<0$ and $n_3\ne1$, $n_5=n_7=1$, we can raise $n_8$ or $n_7$ by~(\ref{dt2});
when $n_8<0$ and $n_2\ne1$, $n_3=n_5=n_7=1$,
we can raise $n_8$ or $n_5$ by~(\ref{dt14});
when $n_8<0$ and $n_2=n_3=n_5=n_7=1$,
we can raise $n_8$ or $n_2$ by~(\ref{dt15}).
When $n_1<0$ and $n_3\ne1$, we can raise $n_1$ by~(\ref{dt13});
when $n_1<0$ and $n_4\ne1$, we can raise $n_1$
using the relation symmetric to~(\ref{dt1});
when $n_1<0$ and $n_2\ne1$, we can raise $n_1$ by~(\ref{dt7});
when $n_1<0$ and $n_7\ne1$, we can raise $n_1$ by~(\ref{dt8});
when $n_1<0$ and $n_8\ne1$, we can raise $n_1$ by~(\ref{dt14});
when $n_1<0$ and $n_2=n_3=n_4=n_7=n_8=1$,
we can raise $n_1$, $n_2$ or $n_3$ by~(\ref{dhom}).
When $n_5<0$ and $n_8\ne1$, we can raise $n_5$ by~(\ref{dt16});
when $n_5<0$ and $n_7\ne1$, we can raise $n_5$ by~(\ref{dt6});
when $n_5<0$ and $n_2\ne1$, we can raise $n_5$ or $n_8$ by~(\ref{dt14});
when $n_5<0$ and $n_2=n_7=n_8=1$, we can raise $n_5$ or $n_2$ by~(\ref{dt7}).
When $n_7<0$ and $n_8\ne1$, we can raise $n_7$ by~(\ref{dt16});
when $n_7<0$ and $n_5\ne1$, we can raise $n_7$ by~(\ref{dt7});
when $n_7<0$ and $n_3\ne1$, $n_5=n_8=1$, we can raise $n_7$ or $n_8$ by~(\ref{dt2});
when $n_7<0$ and $n_1=n_3=n_5=1$, we can raise $n_7$ or $n_3$ by~(\ref{dt6}).
When all the indices are positive,
we can kill one of the lines 1, 5, 7 by~(\ref{dt16}).

Finally, we consider $J_e(n_1,n_2,n_3,n_5,n_6,n_7,n_8)=
I_d(n_1,n_2,n_3,0,n_5,n_6,n_7,n_8)$ (figure~\ref{shrunk}\emph{e}).
This integral vanishes when indices of the following groups of lines
are non-positive: 6, 13, 18, 17, 58, 78, 57, 325, 237.
It becomes trivial if any of the indices is zero.
Applying $(\partial_1+\partial_3)\cdot(k_3-k_2)$ to the integrand of~(\ref{Id}),
we obtain
\begin{equation}
\Bigl[ d-n_1-n_8-2n_7 + n_1\1+\3- + n_8\8+(\5--\7-) \Bigr] J_e = 0\,.
\label{dt17}
\end{equation}
When $n_1<0$, it can be raised using the relation symmetric to~(\ref{dt10}).
When $n_2<0$, it can be raised by~(\ref{dt10}).
When $n_3<0$ and $n_1\ne1$, we can raise $n_3$ by~(\ref{dt4});
when $n_3<0$ and $n_6\ne1$, we can raise $n_3$
using the relation symmetric to~(\ref{dt1});
when $n_3<0$ and $n_2\ne1$, we can raise $n_3$
using the relation symmetric to~(\ref{dt4});
when $n_3<0$ and $n_7\ne1$, we can raise $n_3$ by~(\ref{dt1});
when $n_3<0$ and $n_8\ne1$, we can raise $n_3$ by~(\ref{dt11});
when $n_3<0$ and $n_1=n_2=n_6=n_7=n_8=1$,
we can raise $n_3$, $n_1$ or $n_2$ by~(\ref{dhom}).
When $n_8<0$ and $n_7\ne1$, we can raise $n_8$
using the relation symmetric to~(\ref{dt7});
when $n_8<0$ and $n_5\ne1$, we can raise $n_8$
using the relation symmetric to~(\ref{dt4});
when $n_8<0$ and $n_1\ne1$, $n_5=n_7=1$, we can raise $n_8$ or $n_7$
using the relation symmetric to~(\ref{dt14});
when $n_8<0$ and $n_1=n_5=n_7=1$, we can raise $n_8$ or $n_1$ by~(\ref{dt17}).
When $n_5<0$ and $n_8\ne1$, we can raise $n_5$ by~(\ref{dt17});
when $n_5<0$ and $n_7\ne1$, we can raise $n_5$
using the relation symmetric to~(\ref{dt13});
when $n_5<0$ and $n_1\ne1$, we can raise $n_8$ or $n_7$
using the relation symmetric to~(\ref{dt14});
when $n_5<0$ and $n_1=n_7=n_8=1$,
we can raise $n_5$, $n_7$ or $n_1$ by~(\ref{dt8}).
When $n_7<0$ and $n_8\ne1$, we can raise $n_7$ by~(\ref{dt17});
when $n_7<0$ and $n_5\ne1$, we can raise $n_7$
using the relation symmetric to~(\ref{dt4});
when $n_7<0$ and $n_1\ne1$, we can raise $n_7$ or $n_8$
using the relation symmetric to~(\ref{dt14});
when $n_7<0$ and $n_1=n_5=n_8=1$, we can raise $n_7$ or $n_1$
using the relation symmetric to~(\ref{dt7}).
When all the indices are positive,
we can kill one of the lines 3, 5, 7 by~(\ref{dt17}).

\subsection{Diagrams with four heavy-quark lines}
\label{H3eg}

Let's define (figure~\ref{top}\emph{e})
\begin{equation}
\begin{array}[b]{rclcrclcrcl}
\multicolumn{5}{r}{\displaystyle\int \frac{N_{13}^{n_{13}} N_{23}^{n_{23}} d^d k_1 d^d k_2 d^d k_3}
{D_1^{n_1} D_2^{n_2} D_3^{n_3} D_4^{n_4} D_5^{n_5} D_6^{n_6} D_7^{n_7} D_8^{n_8}}=}
\nonumber\\[9pt]
&=& \multicolumn{9}{l}{ \displaystyle -i \pi^{3d/2} (-2\omega)^{3d+2(n_{13}+n_{23})-2\sum_{i=5}^8 n_i}
I_e(n_1,n_2,n_3,n_4,n_5,n_6,n_7,n_8;n_{13},n_{23})\,,}
\nonumber\\[5pt]
\qquad D_1 &=&\displaystyle  \frac{(k_3+p)\cdot v}{\omega}\,,&\qquad&
D_2 &=&\displaystyle  \frac{(k_1+k_3+p)\cdot v}{\omega}\,,&\qquad&
D_3 &=&\displaystyle  \frac{(k_1+p)\cdot v}{\omega}\,,
\nonumber\\[7pt]
D_4 &=& \displaystyle \frac{(k_2+p)\cdot  v}{\omega}\,,&\qquad&
D_5 &=& -k_3^2\,,&\qquad&
D_6 &=& -k_1^2\,,
\nonumber\\
D_7 &=& -k_2^2\,,&\qquad&
D_8 &=& -(k_1-k_2)^2\,,
\nonumber\\
N_{13} &=& 2k_1\cdot k_3\,,&\qquad&
N_{23} &=& 2k_2\cdot k_3\,.
\end{array}
\label{Ie}
\end{equation}
This integral vanishes when the indices of the following groups of lines
are non-positive: 5, 12, 67, 68, 78, 47, 48, 234, 326, 238.
The heavy-quark denominators are linearly dependent: $D_1-D_2+D_3=1$,
and therefore
\begin{equation}
\left[ 1 - \1- + \2- - \3- \right] I_e = 0\,.
\label{et}
\end{equation}
The cases $n_1\le0$, $n_2\le0$, $n_3\le0$ reduce to $I_c$, $I_d$:
\begin{eqnarray*}
&&I_e(-|n_1|,n_2,n_3,n_4,n_5,n_6,n_7,n_8;n_{13},n_{23}) =
(1+\1--\3-)^{|n_1|} \times\\
&&\times (\6--\4-+\8-)^{n_{13}} (\5--\7-+\8-+\0+)^{n_{23}}
I_c(n_2,n_4,n_3,0,n_7,n_5,n_8,n_6)\,,\\[5pt]
&&I_e(n_1,-|n_2|,n_3,n_4,n_5,n_6,n_7,n_8;n_{13},n_{23}) =\\
&& = (\1-+\3--1)^{|n_2|} (\6--\4--\8-)^{n_{13}} (\0+)^{n_{23}}
I_c(n_1,n_4,n_3,n_5,n_7,0,n_8,n_6)\,,\\[5pt]
&&I_e(n_1,n_2,-|n_3|,n_4,n_5,n_6,n_7,n_8;n_{13},n_{23}) =\\
&& = (1-\1-+\3-)^{|n_3|} (\4--\6-+\8-+\0+)^{n_{13}} (\0+)^{n_{23}}
I_d(n_1,n_4,n_2,n_5,n_7,0,n_6,n_8)\,.
\end{eqnarray*}
If $n_{1,2,3}$ are all positive, we can lower them by~(\ref{et})
until one of them vanish.

Let's define (figure~\ref{top}\emph{f})
\begin{equation}
\begin{array}[b]{rclcrclcrcl}
\multicolumn{7}{r}{\displaystyle \int \frac{N_{13}^{n_{13}} N_{23}^{n_{23}} d^d k_1 d^d k_2 d^d k_3}
{D_1^{n_1} D_2^{n_2} D_3^{n_3} D_4^{n_4} D_5^{n_5} D_6^{n_6} D_7^{n_7} D_8^{n_8}}=\qquad\qquad\qquad}
\nonumber\\[9pt]
&=&\multicolumn{9}{l}{\displaystyle -i \pi^{3d/2} (-2\omega)^{3d+2(n_{13}+n_{23})-2\sum_{i=5}^8 n_i}
I_f(n_1,n_2,n_3,n_4,n_5,n_6,n_7,n_8;n_{13},n_{23})\,,}
\nonumber\\[5pt]
D_1 &=& \displaystyle \frac{(k_1+p)\cdot  v}{\omega}\,,&\qquad&
D_2 &=& \displaystyle \frac{(k_1+k_3+p)\cdot v}{\omega}\,,&\qquad&
D_3 &=& \displaystyle \frac{(k_2+k_3+p)\cdot v}{\omega}\,,
\nonumber\\[7pt]
D_4 &=& \displaystyle  \frac{(k_2+p)\cdot v}{\omega}\,,&\qquad&
D_5 &=& -k_3^2\,,&\qquad&
D_6 &=& -k_1^2\,,
\nonumber\\
D_7 &=& -k_2^2\,,&\qquad&
D_8 &=& -(k_1-k_2)^2\,.
\end{array}
\label{If}
\end{equation}
This integral is mirror-symmetric with respect to $1\leftrightarrow4$,
$2\leftrightarrow3$, $6\leftrightarrow7$.
It vanishes when indices of the following groups of lines
are non-positive: 5, 23, 67, 68, 78, 216, 128, 347, 438.
The heavy-quark denominators are linearly dependent: $D_1-D_2+D_3-D_4=0$,
and therefore
\begin{equation}
\left[ \1- - \2- + \3- - \4- \right] I_f = 0\,.
\label{ft}
\end{equation}
The cases $n_1\le0$, $n_2\le0$ reduce to $I_c$, $I_d$:
\begin{eqnarray*}
&&I_f(-|n_1|,n_2,n_3,n_4,n_5,n_6,n_7,n_8;n_{13},n_{23}) =
(\1-+\2--\3-)^{|n_1|} \times\\
&&\times (\6--\4-+\8-)^{n_{13}} (\5--\7-+\8-+\0+)^{n_{23}}
I_d(n_2,n_4,n_3,0,n_7,n_5,n_8,n_6)\,,\\[5pt]
&&I_f(n_1,-|n_2|,n_3,n_4,n_5,n_6,n_7,n_8;n_{13},n_{23}) =
(\1-+\2--\3-)^{|n_2|} \times\\
&&\times (\4--\6-+\8-+\0+)^{n_{13}} (\7-+\8--\5-)^{n_{23}}
I_c(n_1,n_3,n_4,n_6,0,n_8,n_5,n_7)
\end{eqnarray*}
(the cases $n_4\le0$, $n_3\le0$ are symmetric).
If $n_{1,2,3,4}$ are all positive, we can use~(\ref{ft}) to raise, say, $n_1$
and kill one of the lines 2, 3, 4.

Let's define (figure~\ref{top}\emph{g})
\begin{equation}
\begin{array}[b]{rclcrclcrcl}
\multicolumn{7}{r}{\displaystyle\int \frac{N_{13}^{n_{13}}
N_{23}^{n_{23}} d^d k_1 d^d k_2 d^d k_3} {D_1^{n_1} D_2^{n_2}
D_3^{n_3} D_4^{n_4} D_5^{n_5} D_6^{n_6} D_7^{n_7} D_8^{n_8}}=\qquad\qquad\qquad}
\nonumber\\[7pt]
&=&\multicolumn{9}{l}{\displaystyle -i \pi^{3d/2} 
(-2\omega)^{3d+2(n_{13}+n_{23})-2\sum_{i=4}^8 n_i}
I_g(n_1,n_2,n_3,n_4,n_5,n_6,n_7,n_8;n_{13},n_{23})}
\nonumber\\[3pt]
D_1 &=&\displaystyle \frac{(k_3+p)\cdot  v}{\omega}\,,&\qquad&
D_2 &=&\displaystyle \frac{(k_1+k_3+p)\cdot v}{\omega}\,,&\qquad&
D_3 &=&\displaystyle \frac{(k_2+k_3+p)\cdot v}{\omega}\,,
\nonumber\\[4pt]
D_4 &=&\displaystyle \frac{(k_2+p)\cdot v}{\omega}\,,&\qquad&
D_5 &=& -k_3^2\,,&\qquad&
D_6 &=& -k_1^2\,,
\nonumber\\
D_7 &=& -k_2^2\,,&\qquad&
D_8 &=& -(k_1-k_2)^2\,.
\vspace{-1.5em}\end{array}
\label{Ig}
\end{equation}
This integral vanishes when the indices of the following groups of
lines are non-positive: 5, 67, 68, 78, 26, 28, 123, 234, 347, 438.
The heavy-quark denominators are linearly dependent: $D_1-D_3+D_4=1$,
and therefore
\begin{equation}
\left[ 1 - \1- + \3- - \4- \right] I_g = 0\,.
\label{gt}
\end{equation}
The cases $n_1\le0$, $n_3\le0$, $n_4\le0$ reduce to $I_b$, $I_d$:
\begin{eqnarray*}
&&I_g(-|n_1|,n_2,n_3,n_4,n_5,n_6,n_7,n_8;n_{13},n_{23}) =
(1-\2-+\3-)^{|n_1|} \times\\
&&\times (\6--\4-+\8-)^{n_{13}} (\5--\7-+\8-+\0+)^{n_{23}}
I_d(n_2,n_4,n_3,0,n_7,n_5,n_8,n_6)\,,\\[5pt]
&&I_g(n_1,n_2,-|n_3|,n_4,n_5,n_6,n_7,n_8;n_{13},n_{23}) =
(\1-+\2--1)^{|n_3|} \times\\
&&\times (\4--\6-+\8-+\0+)^{n_{13}} (0+)^{n_{23}}
I_d(n_1,n_4,n_2,n_5,n_7,0,n_6,n_8)\,,\\[5pt]
&&I_g(n_1,n_2,n_3,-|n_4|,n_5,n_6,n_7,n_8;n_{13},n_{23}) =
(1-\1-+\2-)^{|n_4|} \times\\
&&\times (\4-+\6--\0+)^{n_{13}} (\4--\5-+\8-)^{n_{23}}
I_b(n_1,n_3,n_2,n_5,0,n_6,n_8,n_7)\,.
\end{eqnarray*}
If $n_{1,3,4}$ are all positive, we can lower them by~(\ref{gt})
until one of them vanish.

\subsection{Diagrams with five heavy-quark lines}
\label{H3hj}

Let's define (figure~\ref{top}\emph{h})
\begin{equation}
\begin{array}[b]{rclcrclcrcl}
\multicolumn{5}{r}{\displaystyle\int \frac{N_{12}^{n_{12}} N_{13}^{n_{13}} N_{23}^{n_{23}} d^d k_1 d^d k_2 d^d k_3}
{D_1^{n_1} D_2^{n_2} D_3^{n_3} D_4^{n_4} D_5^{n_5} D_6^{n_6} D_7^{n_7} D_8^{n_8}}}
&=& \multicolumn{5}{l}{\displaystyle-i \pi^{3d/2} (-2\omega)^{3d+2(n_{12}+n_{13}+n_{23})-2\sum_{i=6}^8 n_i}\times}
\nonumber\\[2pt]
&&&&&&\multicolumn{5}{l}{\times I_h(n_1,n_2,n_3,n_4,n_5,n_6,n_7,n_8;n_{12},n_{13},n_{23})}
\nonumber\\[9pt]
D_1 &=& \displaystyle \frac{(k_2+p)\cdot v}{\omega}\,,&\qquad&
  D_2 &=& \displaystyle \frac{(k_1+k_2+p)\cdot  v}{\omega}\,,&\qquad&
  D_3 &=& \displaystyle \frac{(k_1+p)\cdot v}{\omega}\,,
\nonumber\\[7pt]
D_4 &=&  \displaystyle \frac{(k_1+k_3+p)\cdot v}{\omega}\,,&\qquad&
  D_5 &=& \displaystyle \frac{(k_3+p)\cdot  v}{\omega}\,,&\qquad&
D_6 &=& -k_2^2\,,
\nonumber\\[7pt]
  D_7 &=& -k_3^2\,,&\qquad&
  D_8 &=& -k_1^2\,,
\nonumber\\[7pt]
N_{12} &=& 2k_1\cdot k_2\,,&\qquad&
  N_{13} &=& 2k_1\cdot k_3\,,&\qquad&
  N_{23} &=& 2k_2\cdot k_3\,.
\end{array}
\label{Ih}
\end{equation}
This integral is mirror-symmetric with respect to $1\leftrightarrow5$,
$2\leftrightarrow4$, $6\leftrightarrow7$.  It vanishes when the
indices of the following groups of lines are non-positive: 6, 7, 8,
12, 45, 234.  There are two linear relations among the heavy-quark
denominators:
\begin{equation}
\left[ 1 - \1- + \2- - \3- \right] I_h = 0\,,\qquad
\left[ 1 - \3- + \4- - \5- \right] I_h = 0\,.
\label{eh}
\end{equation}
The cases
\begin{eqnarray*}
&&I_h(n_1,-|n_2|,n_3,-|n_4|,n_5,n_6,n_7,n_8;n_{12},n_{13},n_{23}) =\\
&& = K_a(n_1,n_5,n_3,n_6,n_7,n_8;|n_2|,|n_4|;n_{23},n_{12},n_{13})\,,\\[2pt]
&&I_h(n_1,-|n_2|,n_3,n_4,-|n_5|,n_6,n_7,n_8;n_{12},n_{13},n_{23}) =\\
&& = K_b(n_1,n_4,n_3,n_6,n_7,n_8;|n_2|,|n_5|,0,0;n_{12},n_{13},n_{23})\,,\\[2pt]
&&I_h(n_1,-|n_2|,-|n_3|,n_4,n_5,n_6,n_7,n_8;n_{12},n_{13},n_{23}) =\\
&& = K_b(n_1,n_4,n_5,n_6,n_8,n_7;0,|n_3|,0,|n_2|;n_{23},n_{13},n_{12})\,,\\[2pt]
&&I_h(-|n_1|,n_2,n_3,n_4,-|n_5|,n_6,n_7,n_8;n_{12},n_{13},n_{23}) =\\
&& = K_c(n_2,n_4,n_3,n_6,n_7,n_8;|n_1|,|n_5|,0;n_{12},n_{13},n_{23})\,,\\[2pt]
&&I_h(-|n_1|,n_2,-|n_3|,n_4,n_5,n_6,n_7,n_8;n_{12},n_{13},n_{23}) =\\
&& = K_e(n_2,n_5,n_4,n_7,n_8,n_6;|n_1|,0,|n_3|;n_{12},n_{23},n_{13})\,,
\end{eqnarray*}
as well as the symmetric cases $n_4\le0$, $n_1\le0$; $n_4\le0$, $n_3\le0$;
$n_5\le0$, $n_3\le0$, reduce to $I_c$, $I_d$,
as will be discussed later in this section.
When $n_3\le0$ (figure~\ref{shrunk}\emph{f}), we can use
\begin{equation}
\left[ \1- - \2- + \4- - \5- \right] I_h = 0
\label{eh2}
\end{equation}
to raise, say, $n_1$ and kill one of the lines 2, 4, 5.
When $n_{1,2}$ are both positive, we can use~(\ref{eh}) to kill one of the lines 1, 2, 3;
when $n_{4,5}$ are both positive, we can kill one of the lines 3, 4, 5.

Let's define (figure~\ref{top}\emph{i})
\begin{equation}
\begin{array}[b]{rclcrclcrcl}
\multicolumn{5}{r}{\displaystyle\int \frac{N_{12}^{n_{12}} 
N_{13}^{n_{13}} N_{23}^{n_{23}} d^d k_1 d^d k_2 d^d k_3}
{D_1^{n_1} D_2^{n_2} D_3^{n_3} D_4^{n_4} D_5^{n_5} D_6^{n_6} D_7^{n_7} D_8^{n_8}}}
&=&\multicolumn{5}{l}{\displaystyle -i \pi^{3d/2} 
(-2\omega)^{3d+2(n_{12}+n_{13}+n_{23})-2\sum_{i=6}^8 n_i}\times}
\nonumber\\
&&&&&&\multicolumn{5}{l}{\times
I_i(n_1,n_2,n_3,n_4,n_5,n_6,n_7,n_8;n_{12},n_{13},n_{23})\,,}
\nonumber\\
 D_1 &=&\displaystyle \frac{(k_1+p)\cdot  v}{\omega}\,,&\quad&
D_2 &=&\displaystyle \frac{(k_1+k_2+p)\cdot v}{\omega}\,,&\quad&
\nonumber\\[7pt]
D_3 &=&\displaystyle \frac{(k_1+k_2+k_3+p)\cdot v}{\omega}\,,&\quad&
  D_4 &=&\displaystyle \frac{(k_1+k_3+p)\cdot v}{\omega}\,,&\quad&
  D_5 &=& \displaystyle \frac{(k_2+p)\cdot v}{\omega}\,,
\nonumber\\
  D_6 &=& -k_1^2\,,&\quad&
  D_7 &=& -k_2^2\,,&\qquad&
  D_8 &=& -k_3^2\,.
\end{array}
\label{Ii}
\end{equation}
This integral vanishes when the indices of the following groups of lines
are non-positive: 6, 7, 8, 34, 123.
There are two linear relations among the heavy-quark denominators:
\begin{equation}
\left[ 1 - \1- + \3- - \4- \right] I_i = 0\,,\qquad
\left[ 1 - \1- + \2- - \5- \right] I_i = 0\,.
\label{ei}
\end{equation}
The cases
\begin{eqnarray*}
&&I_i(n_1,-|n_2|,-|n_3|,n_4,n_5,n_6,n_7,n_8;n_{12},n_{13},n_{23}) =\\
&& = K_b(n_1,n_4,n_5,n_6,n_8,n_7;|n_2|,0,|n_3|,0;n_{12},n_{23},n_{13})\,,\\[2pt]
&&I_i(-|n_1|,n_2,-|n_3|,n_4,n_5,n_6,n_7,n_8;n_{12},n_{13},n_{23}) =\\
&& = K_c(n_2,n_4,n_5,n_6,n_8,n_7;|n_1|,0,|n_3|;n_{12},n_{23},n_{13})\,,\\[2pt]
&&I_i(-|n_1|,-|n_2|,n_3,n_4,n_5,n_6,n_7,n_8;n_{12},n_{13},n_{23}) =\\
&& = K_d(n_3,n_5,n_4,n_7,n_6,n_8;|n_1|,0,0,|n_2|;n_{23},n_{12},n_{13})\,,\\[2pt]
&&I_i(-|n_1|,n_2,n_3,-|n_4|,n_5,n_6,n_7,n_8;n_{12},n_{13},n_{23}) =\\
&& = K_d(n_3,n_5,n_2,n_7,n_8,n_6;0,|n_1|,0,|n_4|;n_{12},n_{23},n_{13})\,,\\[2pt]
&&I_i(n_1,n_2,n_3,-|n_4|,-|n_5|,n_6,n_7,n_8;n_{12},n_{13},n_{23}) =\\
&& = K_d(n_3,n_1,n_2,n_6,n_8,n_7;0,|n_5|,|n_4|,0;n_{12},n_{13},n_{23})\,,\\[2pt]
&&I_i(n_1,n_2,-|n_3|,n_4,-|n_5|,n_6,n_7,n_8;n_{12},n_{13},n_{23}) =\\
&& = K_e(n_4,n_1,n_2,n_6,n_7,n_8;0,|n_3|,|n_5|;n_{23},n_{13},n_{12})\,,\\[2pt]
&&I_i(n_1,-|n_2|,n_3,-|n_4|,n_5,n_6,n_7,n_8;n_{12},n_{13},n_{23}) =\\
&& = K_f(n_1,n_5,n_3,n_6,n_7,n_8;|n_2|,0,|n_4|;n_{13},n_{23},n_{12})\,,\\[2pt]
&&I_i(-|n_1|,n_2,n_3,n_4,-|n_5|,n_6,n_7,n_8;n_{12},n_{13},n_{23}) =\\
&& = K_g(n_2,n_4,n_3,n_6,n_8,n_7;|n_5|,|n_1|,0;n_{12},n_{23},n_{13})
\end{eqnarray*}
reduce to $I_c$, $I_d$, as will be discussed later in this section.
When $n_1\le0$, we can use
\begin{equation}
\left[ \2- - \3- + \4- - \5- \right] I_i = 0
\label{ei2}
\end{equation}
to raise, say, $n_2$ and kill one of the lines 3, 4, 5.
When $n_{2,5}$ are both positive, we can use~(\ref{ei}) to kill one of the lines 1, 2, 5;
when $n_{3,4}$ are both positive, we can kill one of the lines 1, 3, 4.

Finally, let's define (figure~\ref{top}\emph{j})
\begin{equation}
\begin{array}[b]{rclcrclcrcl}
\multicolumn{5}{r}{\displaystyle\int \frac{N_{12}^{n_{12}}
N_{13}^{n_{13}} N_{23}^{n_{23}} d^d k_1 d^d k_2 d^d k_3} {D_1^{n_1}
D_2^{n_2} D_3^{n_3} D_4^{n_4} D_5^{n_5} D_6^{n_6} D_7^{n_7} D_8^{n_8}}}
&=& \multicolumn{5}{l}{-i \pi^{3d/2}
(-2\omega)^{3d+2(n_{12}+n_{13}+n_{23})-2\sum_{i=6}^8 n_i}\times}
\nonumber\\[7pt]
&&&&&&\multicolumn{5}{l}{\times I_j(n_1,n_2,n_3,n_4,n_5,n_6,
n_7,n_8;n_{12},n_{13},n_{23})\,,}
\nonumber\\
D_1 &=&  \displaystyle\frac{(k_1+p)\cdot v}{\omega}\,,&\quad&
  D_2 &=&  \displaystyle\frac{(k_1+k_2+p)\cdot v}{\omega}\,,
\nonumber\\[7pt]
D_3 &=&  \displaystyle\frac{(k_1+k_2+k_3+p)\cdot v}{\omega}\,,&\quad&
  D_4 &=&  \displaystyle\frac{(k_2+k_3+p)\cdot v}{\omega}\,,&\quad&
D_5 &=&  \displaystyle \frac{(k_3+p)\cdot v}{\omega}\,,
\nonumber\\
  D_6 &=& -k_1^2\,,&\quad&
  D_7 &=& -k_2^2\,,&\quad&
  D_8 &=& -k_3^2\,.
\end{array}
\label{Ij}
\end{equation}
This integral is mirror-symmetric with respect to $1\leftrightarrow5$,
$2\leftrightarrow4$, $6\leftrightarrow8$.  It vanishes when the
indices of the following groups of lines are non-positive: 6, 7, 8,
123, 234, 345.
There are two linear relations among the heavy-quark denominators:
\begin{equation}
\left[ 1 - \1- + \3- - \4- \right] I_j = 0\,,\qquad
\left[ 1 - \2- + \3- - \5- \right] I_j = 0\,.
\label{ej}
\end{equation}
The cases
\begin{eqnarray*}
&&I_j(n_1,-|n_2|,-|n_3|,n_4,n_5,n_6,n_7,n_8;n_{12},n_{13},n_{23}) =\\
&& = K_b(n_1,n_4,n_5,n_6,n_7,n_8;0,0,|n_3|,|n_2|;n_{13},n_{23},n_{12})\,,\\[2pt]
&&I_j(-|n_1|,-|n_2|,n_3,n_4,n_5,n_6,n_7,n_8;n_{12},n_{13},n_{23}) =\\
&& = K_d(n_3,n_5,n_4,n_8,n_6,n_7;|n_1|,0,|n_2|,0;n_{23},n_{13},n_{12})\,,\\[2pt]
&&I_j(-|n_1|,n_2,-|n_3|,n_4,n_5,n_6,n_7,n_8;n_{12},n_{13},n_{23}) =\\
&& = K_e(n_2,n_5,n_4,n_8,n_7,n_6;|n_1|,|n_3|,0;n_{12},n_{13},n_{23})\,,\\[2pt]
&&I_j(n_1,-|n_2|,n_3,-|n_4|,n_5,n_6,n_7,n_8;n_{12},n_{13},n_{23}) =\\
&& = K_f(n_1,n_5,n_3,n_6,n_8,n_7;0,|n_2|,|n_4|;n_{12},n_{23},n_{13})\,,\\[2pt]
&&I_j(-|n_1|,n_2,n_3,n_4,-|n_5|,n_6,n_7,n_8;n_{12},n_{13},n_{23}) =\\
&& = K_g(n_2,n_4,n_3,n_6,n_8,n_7;0,|n_1|,|n_5|;n_{12},n_{23},n_{13})\,,
\end{eqnarray*}
as well as the symmetric cases $n_4\le0$, $n_3\le0$; $n_5\le0$, $n_4\le0$;
$n_5\le0$, $n_3\le0$, reduce to $I_c$, $I_d$,
as will be discussed later in this section.
When $n_3\le0$, we can use
\begin{equation}
\left[ \1- - \2- + \4- - \5- \right] I_j = 0
\label{ej2}
\end{equation}
to raise, say, $n_1$ and kill one of the lines 2, 4, 5.
When $n_{1,4}$ are both positive, we can use~(\ref{ej}) to kill one of the lines 1, 3, 4;
when $n_{2,5}$ are both positive, we can kill one of the lines 2, 3, 5.

\FIGURE[t]{
\begin{picture}(132,82)
\put(66,41){\makebox(0,0){\epsfig{file=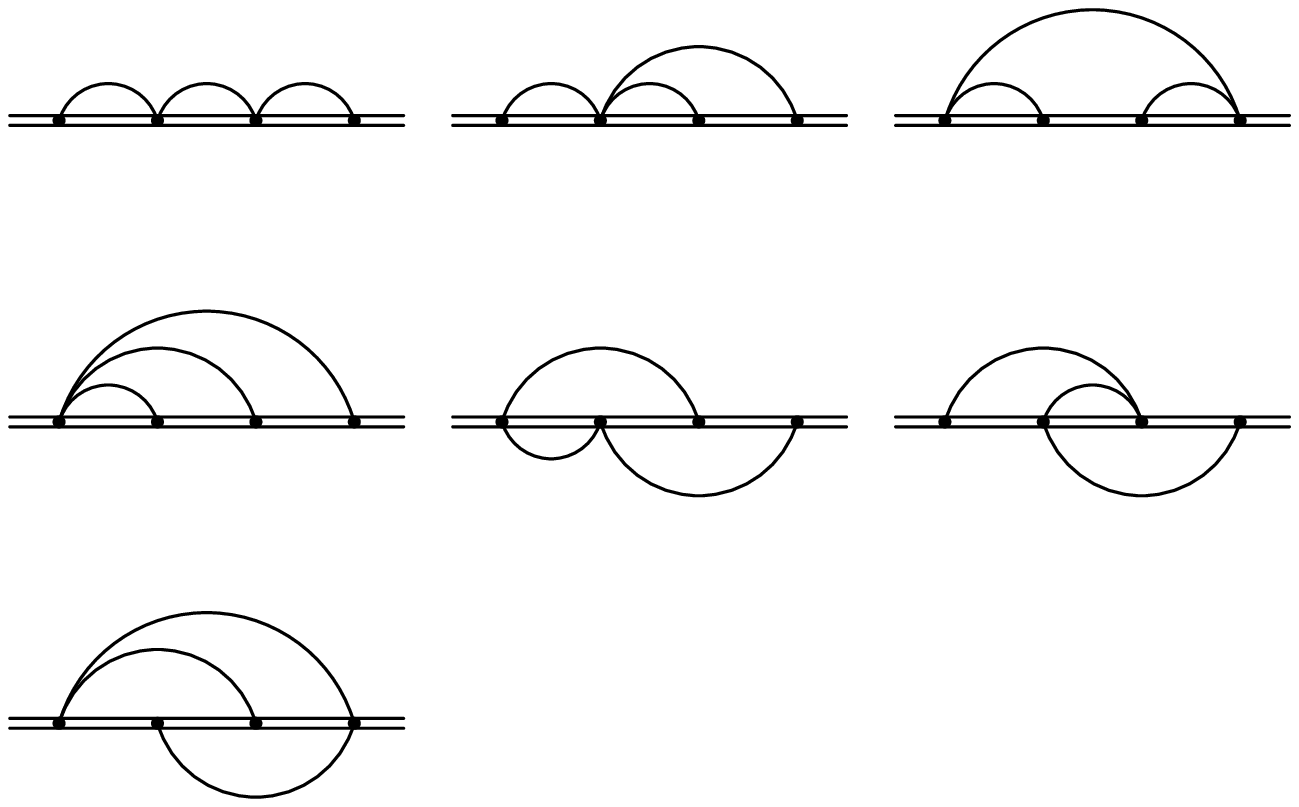}}}
\put(21,58.25){\makebox(0,0)[b]{$a$}}
\put(66,58.25){\makebox(0,0)[b]{$b$}}
\put(111,58.25){\makebox(0,0)[b]{$c$}}
\put(21,27.625){\makebox(0,0)[b]{$d$}}
\put(66,27.625){\makebox(0,0)[b]{$e$}}
\put(111,27.625){\makebox(0,0)[b]{$f$}}
\put(21,-3){\makebox(0,0)[b]{$g$}}
\end{picture}
\caption{Reduced forms of the diagrams figure~\protect\ref{top}\emph{h}--\emph{j}}
\label{red}}

The reduced forms of the integrals $I_{h,i,j}$ are (figure~\ref{red})
\begin{eqnarray*}
&&K_a(n_1,n_2,n_3,n_4,n_5,n_8;n_9,n_{10};n_{45},n_{48},n_{58}) =
(\1-+\3--1)^{n_9} (\2-+\3--1)^{n_{10}} \times\\
&&\quad{}\times (\0+)^{n_{45}} (\6--\4--\8-)^{n_{48}} (\7--\5--\8-)^{n_{58}}
I_c(n_1,n_2,n_3,n_4,n_5,0,0,n_8)\,,\\[4pt]
&&K_b(n_1,n_2,n_3,n_4,n_7,n_8;n_9,n_{10},n_{11},n_{12};n_{48},n_{78},n_{47}) =
(\1-+\3--1)^{n_9} \times\\
&&\quad{}\times (\2--\3-+1)^{n_{10}} (\1-+\2--1)^{n_{11}} (\1-+\2--\3-)^{n_{12}}
(\6--\4--\8-)^{n_{48}} \times\\
&&\quad{}\times (\7--\5-+\8-)^{n_{78}} (\4--\6-+\8-+\0+)^{n_{47}}
I_c(n_1,n_2,n_3,n_4,0,0,n_7,n_8)\,,\\
&&K_c(n_1,n_2,n_3,n_6,n_7,n_8;n_9,n_{10},n_{11};n_{68},n_{78},n_{67}) =
(\1--\3-+1)^{n_9} \times\\
&&\quad{}\times (\2--\3-+1)^{n_{10}} (\1-+\2--\3-)^{n_{11}}
(\6--\4-+\8-)^{n_{68}} (\7--\5-+\8-)^{n_{78}} \times\\
&&\quad{}\times (\4-+\5--\6--\7-+\0+)^{n_{67}}
I_c(n_1,n_2,n_3,0,0,n_6,n_7,n_8)\,,\\
&&K_d(n_1,n_2,n_3,n_5,n_6,n_7;n_9,n_{10},n_{11},n_{12};n_{57},n_{56},n_{67}) =
(\1--\3-+1)^{n_9} \times\\
&&\quad{}\times (\3--\2-+1)^{n_{10}} (\1--\2-+1)^{n_{11}} (\1-+\2--\3-)^{n_{12}}
(\5-+\7--\8-)^{n_{57}} \times\\
&&\quad{}\times (\5--\7-+\8-+\0+)^{n_{56}}
(\6-+\7--\4--\5--\0+)^{n_{67}} \times\\
&&\quad{}\times I_c(n_1,n_2,n_3,0,n_5,n_6,n_7,0)\,,\\
&&K_e(n_1,n_2,n_3,n_5,n_6,n_8;n_9,n_{10},n_{11};n_{68},n_{58},n_{56}) =
(\1-+\2--\3-)^{n_9} \times\\
&&\quad{}\times (\1-+\2--1)^{n_{10}} (\3--\2-+1)^{n_{11}} (\6--\4-+\8-)^{n_{68}}
(\7--\5--\8-)^{n_{58}} \times\\
&&\quad{}\times (\5--\7-+\8-+\0+)^{n_{56}}
I_d(n_1,n_2,n_3,0,n_5,n_6,0,n_8)\,,\\[4pt]
&&K_f(n_1,n_2,n_3,n_4,n_5,n_8;n_9,n_{10},n_{11};n_{48},n_{58},n_{45}) =
(\1-+\2--1)^{n_9} \times\\
&&\quad{}\times (\3--\2-+1)^{n_{10}} (\3--\1-+1)^{n_{11}}
(\4--\6-+\8-)^{n_{48}} (\5--\7-+\8-)^{n_{58}} \times\\
&&\quad{}\times (\0+)^{n_{45}}
I_d(n_1,n_2,n_3,n_4,n_5,0,0,n_8)\,,\\[4pt]
&&K_g(n_1,n_2,n_3,n_6,n_7,n_8;n_9,n_{10},n_{11};n_{68},n_{78},n_{67}) =
(\1-+\2--\3-)^{n_9} \times\\
&&\quad{}\times (\3--\2-+1)^{n_{10}} (\3--\1-+1)^{n_{11}} (\6--\4-+\8-)^{n_{68}}
(\7--\5-+\8-)^{n_{78}} \times\\
&&\quad{}\times (\4-+\5--\6--\7-+\0+)^{n_{67}}
I_d(n_1,n_2,n_3,0,0,n_6,n_7,n_8)\,.
\end{eqnarray*}

\section{Implementation and testing}
\label{Imp}

The package \textsf{Grinder} consists of a set of mutually recursive
procedures for Feynman integrals of various topologies, which reduce a
given Feynman integral to simpler ones, until boundary-case integrals
with known values are reached.
The package is written in \textsf{REDUCE}~\cite{H,G}.
Remembering results of previous
function calls may make the program run much faster, if there is
enough memory
(unfortunately, \textsf{REDUCE} uses linear look-up in remember-tables).

I also re-implemented it in \textsf{Axiom}~\cite{JS}.  All expressions
involved are linear combinations of basis integrals with coefficients
which are rational functions of $d$.  It is convenient to use
\textsf{Axiom} domain \texttt{Vector Fraction UnivariatePolynomial},
which has all the necessary
operations.  This makes intermediate expressions shorter than in the
case when multivariate rational functions are used for entire
expressions, because, typically, not all basis integrals are
accompanied by every possible denominator.  The amount of GCD
calculations is thus reduced.  This improvement can be, in principle,
back-propagated to the \textsf{REDUCE} implementation by using
matrices.  However, working with matrices in \textsf{REDUCE} is
awkward, because there are no local matrix variables, and no easy way
for a function to return a matrix.
On the other hand,
complete diagram calculations, including tensor and $\gamma$-matrix
algebra, can be done in \textsf{REDUCE}.

The main method of testing was checking various recurrence relations
(including those which were not directly used for construction of the
algorithm) in nested loops over $n_i$.  For each of two-loop and
generalized two-loop integrals, which depend on 5 indices, a typical
number of checks was about 20000; each test set of this size runs for
a few hours.  Integrals with 8 or 9 indices are more difficult to
check.  Some test sets required a few days of CPU time.  A tool
showing how many times each linear code segment has been executed
would be invaluable for setting up test cases which check all branches
at least once.  Unfortunately, such a tool is not available in either
programming system, and I had to emulate it by hand.

\acknowledgments

I am grateful to P.A.~Baikov, D.J.~Broadhurst, K.G.~Chetyrkin,
S.A.~Larin and J.A.M.~Vermaseren for useful discussions, and to
INTAS for the grant which allowed me to buy \textsf{Axiom}.

\end{document}